\documentclass[aps,prd,twocolumn,superscriptaddress,showpacs,nofootinbib] {revtex4-1}

\usepackage{epsfig}
\usepackage{graphicx}
\usepackage{dcolumn}
\usepackage{bm}
\usepackage{multirow}
\usepackage{array}
\usepackage{slashed}
\usepackage{amsmath}
\usepackage[titletoc]{appendix}
\usepackage{color}
\usepackage{hyperref}
\usepackage{soul}

\begin{document}

\title{Binary Neutron Star Mergers as Potential Sources for Ultra-High-Energy Cosmic Rays and High-Energy Neutrinos} 

\author{Gang Guo}
\email{guogang@cug.edu.cn} 
\affiliation{School of Mathematics and Physics, China University of Geosciences, Wuhan 430074, China}
\affiliation{Shenzhen Research Institute of China University of Geosciences, Shenzhen 518057, China}
\author{Yong-Zhong Qian}
\email{qian@physics.umn.edu} 
\affiliation{School of Physics and Astronomy, University of Minnesota, Minneapolis, MN 55455, USA}
\author{Meng-Ru Wu}
\email{mwu@as.edu.tw}
\affiliation{Institute of Physics, Academia Sinica, Taipei, 11529, Taiwan}
\affiliation{Institute of Astronomy and Astrophysics, Academia Sinica, Taipei, 10617, Taiwan}
\affiliation{Physics Division, National Center for Theoretical Sciences, Taipei 10617, Taiwan}

\date{\today}

\begin{abstract}

Recent studies suggest that the most energetic cosmic rays, exceeding 100 EeV, may primarily consist of $r$-process nuclei. This highlights binary neutron star mergers and collapsars as promising sources of ultra-high-energy cosmic rays (UHECRs). Building on these insights, we examine the conditions that facilitate the efficient production of UHE $r$-process nuclei during the prompt radiation (PR), extended emission (EE), and plateau emission phases of short gamma-ray bursts (sGRBs) following neutron star mergers. Our study reveals that jets associated with the PR phase, characterized by typical bulk Lorentz factors ($\gtrsim 400$--500), dissipation radii, and magnetic field strengths,
can accelerate $r$-process nuclei to energies $\gtrsim 100$ EeV while preserving them during propagation within the source. Additionally, we investigate the production of HE neutrinos from photomeson and hadronic interactions, as well as from the $\beta$ decay of accelerated $r$-process nuclei. We find that the HE neutrino fluxes from sGRBs, mainly produced via photomeson interactions, are significantly limited to preserve the accelerated heavy nuclei, leading to lower fluxes than the predictions without allowing for contributions to UHECRs. Our results suggest that sGRBs may potentially contribute to UHECRs during the PR phase and to HE neutrinos during the EE phase—a scenario that can be tested by future neutrino observatories.
\end{abstract}


\maketitle

\section{introduction}

Astrophysical objects such as active galactic nuclei (AGN), gamma-ray bursts (GRBs), magnetars, and fast radio bursts (FRBs), have been proposed as potential sources of ultra-high-energy cosmic rays (UHECRs) with energies exceeding $10^{18}$ eV (EeV). However, the exact origin of UHECRs remains an unsolved mystery. Over the past few decades, advances in both observational techniques and theoretical frameworks have greatly reshaped our understanding of UHECRs. For instance, more and more CR events with energies surpassing 100 EeV have been observed by the Telescope Array (TA) \cite{Abu_Zayyad_2013,Ivanov:2020rqn} and the Pierre Auger Observatory \cite{Abraham_2010,PierreAuger:2022qcg}. Notably, the TA recently recorded an extremely energetic CR event of $\sim 244$ EeV (dubbed the `Amaterasu' particle) \cite{TA:2023sbd}, comparable to the famous `Oh-My-God' particle detected by the Fly's Eye experiment three decades ago with an estimated energy of $\sim 300$ EeV. These extraordinarily energetic events challenge our understanding of CR acceleration and propagation, highlighting the need for further exploration and study.

The composition of UHECRs offers critical insights into their origins and has been extensively investigated by Auger and TA. The Auger data suggest a gradual transition in UHECR composition from lighter to heavier nuclei as energy increases, with heavier nuclei becoming more dominant above 10 EeV \cite{Auger:2016use,Auger:2022atd}. Although the TA results are not entirely consistent with those of Auger, they also show a significant transition toward heavier elements at energies exceeding 10 EeV \cite{TelescopeArray:2024buq}. Interestingly, to fit the most energetic TA events above 100 EeV, it appears necessary to consider the presence of heavy nuclei even beyond iron under standard astrophysical assumptions \cite{TelescopeArray:2024buq,TelescopeArray:2024oux}.

Building on these findings and recognizing that particle acceleration primarily depends on the rigidity of charged particles [$\eta \equiv E/(Ze)$], with the maximum achievable $\eta$ being largely independent of particle species, Refs.~\cite{Farrar:2024zsm,Zhang:2024sjp} recently proposed that nuclei with atomic numbers $Z \gtrsim 50$ produced via rapid neutron capture, the $r$-process, could significantly contribute to UHECRs, particularly at the highest observed energies.
If this hypothesis holds true, the sources of UHECRs would also act as sites for $r$-process nucleosynthesis. The optical/infrared transient AT2017gfo, a kilonova associated with GW170817, provided strong evidence that binary neutron star mergers (BNSMs) are an $r$-process site \cite{LIGOScientific:2017vwq, LIGOScientific:2017ync, Kasen:2017sxr, Shibata:2017xdx, Metzger:2019zeh, Watson:2019xjv} (see also \cite{Tanvir:2013pia,Yang:2015pha,Jin:2016pnm}). Additionally, the detection of a short GRB (sGRB), GRB170817A, following the GW170817 event, provided the first direct evidence that particle acceleration within relativistic jets by shocks or magnetic reconnection in BNSMs could produce sGRBs \cite{LIGOScientific:2017vwq, LIGOScientific:2017ync,LIGOScientific:2017zic}. A fraction of the freshly-synthesized $r$-process nuclei in BNSMs can be swept into the relativistic jets and subsequently accelerated, potentially
contributing to the most energetic CR events. Similarly, collapsars thought to generate long GRBs (lGRBs), may also provide neutron-rich conditions suitable for $r$-process nucleosynthesis \cite{Siegel:2018zxq,Cowan:2019pkx,Brauer:2020hty,Fraser:2021piw,Lee:2022ijg,Barnes:2022dxv,Anand:2023ujd,Barnes:2023cpj,Li:2023xur,Dean:2024pne,Fischer:2023ebq}. Therefore, these two types of explosive events are promising sites for producing UHE $r$-process nuclei. Considering the energy generation rates required to account for the observed UHECRs, Ref.~\cite{Zhang:2024sjp} suggested both BNSMs and collapsars as plausible sources for the most energetic CR events above $\sim 100$ EeV, whereas Ref.~\cite{Farrar:2024zsm} focused exclusively on BNSMs.

Motivated by these studies, we aim to investigate in detail the jet conditions in GRBs that enable the acceleration of UHE $r$-process nuclei while ensuring their survival during propagation inside the sources. Following Refs.~\cite{Farrar:2024zsm,Zhang:2024sjp}, we focus primarily on the most energetic CR events with energies above 100 EeV, which are hypothesized to be associated with $r$-process nuclei. While our analysis specifically targets sGRBs originating from BNSMs, similar studies could also be applied to lGRBs. The light curves of sGRBs typically exhibit a prompt radiation (PR) phase,
followed by two distinct components with varying luminosities and durations: an extended emission
(EE) phase and a plateau emission (PE) phase \cite{Kisaka:2015mza,Kisaka:2017tas}. The total energy of the EE phase could be comparable to that of the PR phase, while the PE phase has a total energy typically 1--2 orders of magnitude lower \cite{Kisaka:2017tas,Kimura:2017kan}. We examine the conditions associated with all three phases, focusing on the acceleration of UHE $r$-process heavy nuclei and their survival within the sources. Based on this examination, we obtain important
constraints on the relevant jet parameters, which could provide insights into the potential origins of UHECRs.     

HE neutrinos can be generated through interactions between accelerated nuclei and stellar matter or photons in GRBs \cite{Waxman:1997ti,Waxman:1999ai,Meszaros2001,Li2002,Guetta:2003wi,Murase:2006dr,Murase:2006mm,Hummer:2011ms,Li_2012,Zhang:2012qy,Liu2013,Tamborra2015,Bustamante2015,Kimura:2022zyg}. In addition, they can also be produced from decay of accelerated $r$-process nuclei (see also \cite{Chen:2023skn,An:2023edd} for studies on MeV-scale neutrino emission from decay of $r$-process nuclei in BNSMs). Although stacked searches for HE neutrinos from sGRBs and targeted searches, such as those for GRB170817A, have so far yielded null results \cite{Casier:2015ads,ANTARES:2017bia,Baikal-GVD:2018cya}, detection of HE neutrino signals from sGRBs remains promising within various models \cite{Fang:2017tla,Kimura:2017kan,Biehl:2017qen,Kimura:2018vvz,Decoene:2019eux,Matsui:2023ohr,Mukhopadhyay:2023niv,Mukhopadhyay:2024lwq,Rossoni:2024ial}. In this study, we compute HE neutrino signals from sGRBs during each emission phase, adopting jet parameters suitable for the production of UHE $r$-process nuclei. We find that the resulting HE neutrino signals are more constrained compared to those using typical jet parameters assumed in previous studies.      

\begin{figure*}[htbp]  
\centering
\includegraphics[width=1.0\textwidth]{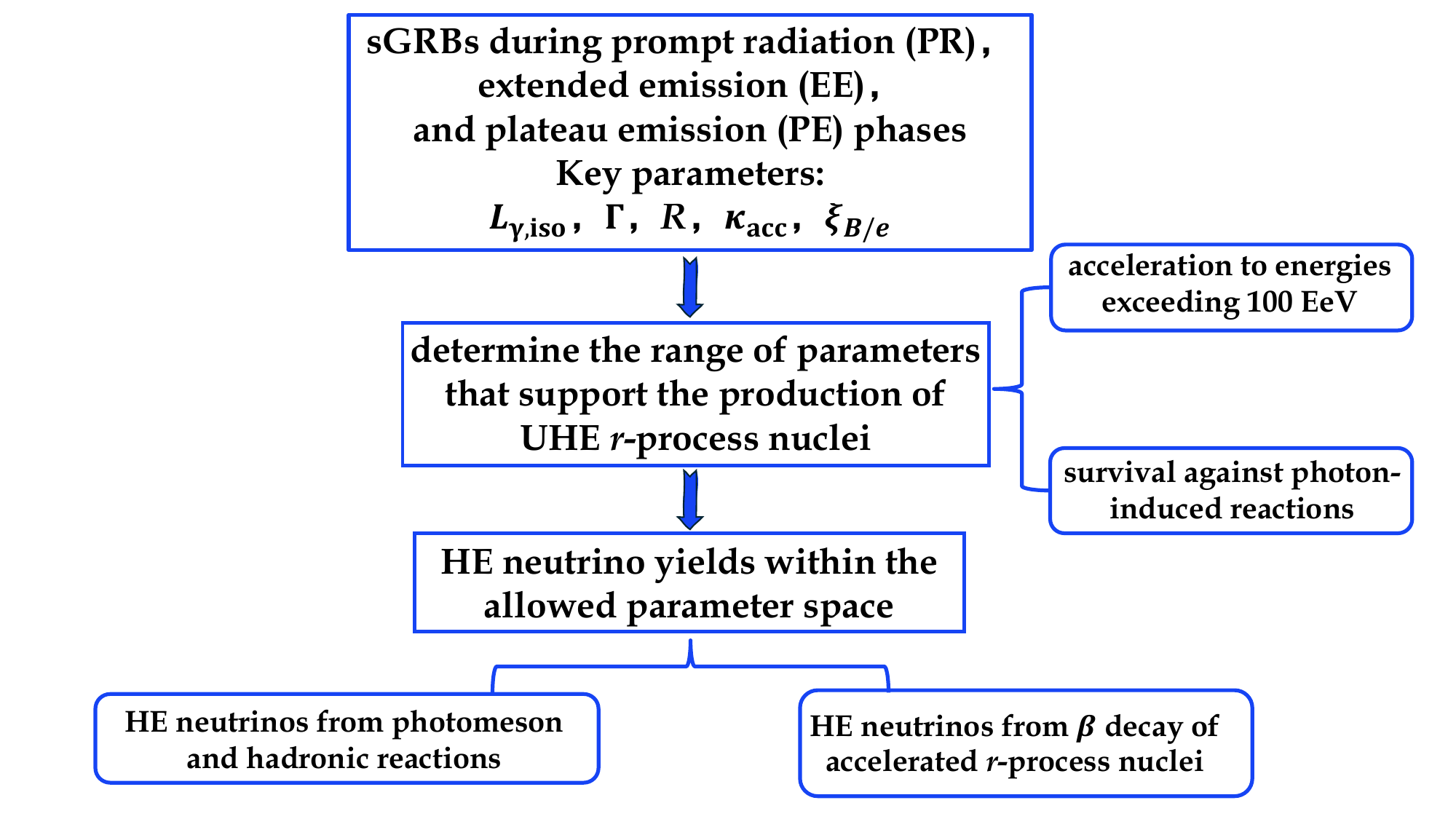} 
\caption{The flowchart of this work.} 
\label{fig:flowchart}  
\end{figure*}

This paper is organized as follows. In Sec.~\ref{sec:timescale}, we calculate the timescales for reactions between heavy nuclei and photons during different emission phases of sGRBs. By evaluating the maximal energies achievable for accelerated heavy nuclei and their survival probabilities at each dissipation site, we derive the constraints on jet parameters necessary for efficient production of the most energetic CR events above 100 EeV in Sec.~\ref{sec:Emax-survival}. In Sec.~\ref{sec:neutrinos}, we study the HE neutrino fluxes from each phase and discuss the potential of using these neutrino signals to better understand the origins of UHECRs. We give conclusions in Sec.~\ref{sec:summary}. For clarity, a flowchart outlining this work is presented in Fig.~\ref{fig:flowchart}.
The term ``heavy nuclei" refers specifically to $r$-process nuclei unless noted otherwise.

\section{photon density from SGRBS and timescales for $A\gamma$ reactions}
\label{sec:timescale}

We consider three different dissipation regions within an sGRB associated with the prompt radiation (PR), the extended emission (EE), and the plateau emission (PE), respectively. The energy-differential number density of photons for each dissipation region in the fluid rest frame can be characterized by a broken power-law distribution as \cite{Kimura:2017kan}
\begin{align}
n(E_\gamma) = n_0 \times \left\{ 
\begin{array}{cc}
\Big({E_\gamma \over E_{\gamma b}}\Big)^{-\alpha},& E_\gamma \le E_{\gamma b}, \\
\Big({E_\gamma \over E_{\gamma b}}\Big)^{-\beta},& E_\gamma > E_{\gamma b}. 
\end{array}
\right.
\label{eq:n_gamma}
\end{align}
Here, $E_\gamma$ is photon energy in the fluid rest frame, $E_{\gamma b}$ is the break energy, $\alpha$ and $\beta$ are the power-law indices, and $n_0$ is a normalization constant. The minimal and maximal values of $E_\gamma$ are set to $E_\gamma^{\rm min}=0.1$ eV and $E_\gamma^{\rm max}=1$ MeV. Unless specified otherwise, we denote $E_i$ as the energy of species $i$ in the fluid frame and use $E_i^{\rm ob} \approx \Gamma E_i$ for the observed energy in the laboratory frame, where $\Gamma$ is the bulk Lorentz factor of the fluid.\footnote{For simplicity, we neglect the effect of the cosmological redshift on the observed energy.}
Assuming steady photon luminosity during each emission phase, the normalization constant $n_0$ is determined by $\int_{E_\gamma^{\rm min}}^{E_\gamma^{\rm max}} n(E_\gamma) E_\gamma dE_\gamma = L_{\gamma, \rm iso}/(4\pi \Gamma^2 R^2 c)$, where $L_{\gamma, \rm iso}$ is the photon luminosity integrated over the allowed energy bands, $R$ is the dissipation radius, and
$c$ is the speed of light. We adopt the following typical isotropic photon luminosities for the emission phases: $L_{\gamma, \rm iso}=10^{52}$ erg/s for PR \cite{Wanderman:2014eza,Shahmoradi:2014ira,Ghirlanda:2016ijf}, $10^{49}$ erg/s for EE \cite{Kagawa:2015gaa,Kaneko:2015lga,Kisaka:2017tas}, and $10^{47}$ erg/s for PE \cite{Evans:2008wp,Rowlinson:2013ue,Kisaka:2017tas}, noting that these values can vary within a relatively broad range. The typical bulk Lorentz factor for PR lies between $\sim 10^2$ and $\sim 10^3$ \citep{piran2005,Meszaros2006,Kumar2014}, whereas for EE and PE, it ranges from $\sim 10$ to $\sim 10^2$ \cite{Matsumoto:2020fle}. The dissipation radii $R$ depend on specific models and vary between $10^{12}$ and $10^{17}$~cm for all emission phases in our study. For the EE phase, we consider two scenarios following Ref.~\cite{Kimura:2017kan}: one (EE-H) with $E_{\gamma b}^{\rm ob}=10$ keV and the other (EE-L) with $E_{\gamma b}^{\rm ob}=1$ keV. For simplicity, we fix $\alpha=0.5$ and $\beta=2$. Table~\ref{tab:pars} summarizes the values of $L_{\gamma, \rm iso}$, $E_{\gamma b}^{\rm ob}$, and the ranges of $\Gamma$ and $R$ considered for each emission phase. All chosen parameters are broadly consistent with those used in Ref.~\cite{Kimura:2017kan}.

\begin{table*}[htbp] 
\centering 
\renewcommand{\arraystretch}{1.5}
\caption{Relevant parameters and their typical values considered in this study for the prompt radiation (PR), extended emission (EE), and plateau emission (PE): the isotropic photon luminosity $L_{\gamma, \rm iso}$ integrated over allowed bands, the total energy of  emitted photons $\mathcal{E}_{\gamma, \rm iso}$, the observed peak photon energy $E_{\gamma b}^{\rm ob}$ at the spectral break, the fluid bulk Lorentz factor $\Gamma$, the dissipation radius $R$, the acceleration efficiency parameter $\kappa_{\rm acc}$, and the energy ratio of magnetic field to electron $\xi_{B/e}$. The minimal and maximal photon energies in the fluid frame are set to $E^{\rm min}_\gamma=0.1$ eV and $E^{\rm max}_\gamma=1$ MeV, and the photon spectral indices are set to be $\alpha=0.5$ and $\beta=2$ below and above $E_{\gamma b}$, respectively [Eq.~\eqref{eq:n_gamma}].
\label{tab:pars}}
\begin{ruledtabular}
\begin{tabular}{ccccccccc}    
Parameters & $L_{\gamma, \rm iso}$~(erg/s) &
$\mathcal{E}_{\gamma, \rm iso}$ (erg) &
$E_{\gamma b}^{\rm ob}$~(keV) & $\Gamma$ & $R$ (cm) & $\kappa_{\rm acc}$ & $\xi_{B/e}$  \\    
\hline
PR & $10^{52}$ & $10^{52}$ &  500 & 100--1000 & $10^{12}$--$10^{17}$ & 1--100 & $10^{-3}$--$10^{3}$  \\ 
EE-H & $10^{49}$ & $3 \times 10^{51}$ & 10 & 10--100 & $10^{12}$--$10^{17}$ & 1--100 & $10^{-3}$--$10^{3}$  \\ 
EE-L & $10^{49}$ & $3 \times 10^{51}$ &  1 & 10--100 & $10^{12}$--$10^{17}$ & 1--100 & $10^{-3}$--$10^{3}$ \\ 
PE & $10^{47}$ & $3 \times 10^{50}$ & 0.1 & 10--50 & $10^{12}$--$10^{17}$ & 1--100 & $10^{-3}$--$10^{3}$  \\ 
\end{tabular}
\end{ruledtabular}
\end{table*}

Different types of nucleus-photon ($A\gamma$) reactions dominate at different energies. At $\epsilon_\gamma > 2m_e$ with $\epsilon_\gamma$ being the photon energy in the rest frame of nuclei, $e^\pm$ pairs can be produced via the Bethe-Heitler (BH) process. 
When $\epsilon_\gamma$ becomes comparable with the nuclear binding energy per nucleon, i.e., $\epsilon_\gamma \gtrsim 10$ MeV, photodisintegration of nuclei becomes possible. For $\epsilon_\gamma \gtrsim 0.14$ GeV, pion production dominates.      

The reaction timescales $t_{{\rm reac}, i}(E_A)$ for the $A\gamma$ reactions in the fluid rest frame can be expressed as 
\begin{align}
t^{-1}_{{\rm reac}, i}(E_A) =& {c\over 2\Gamma_A^2} \int_{\epsilon_{{\rm
th},i}}^{2\Gamma_AE_\gamma^{\rm max}} d\epsilon_\gamma \epsilon_\gamma \sigma_i(\epsilon_\gamma) \nonumber \\ 
& \times \int^{E_\gamma^{\rm max}}_{\max[E_\gamma^{\rm min},\epsilon_\gamma/(2\Gamma_A)]} dE_\gamma E_\gamma^{-2} n(E_\gamma), 
\label{eq:trec}
\end{align}
where $i=$ `BH', `photodis', or `meson' for the BH process, photodisintegration, or meson production, $E_A$ and $E_\gamma$ are the nucleus and photon energy, respectively, in the fluid rest frame,
$\Gamma_A = E_A/(m_Ac^2)$ with $m_A$ being the nucleus mass, 
$\epsilon_{{\rm th}, i}$ is the threshold energy
of $\epsilon_\gamma$, and $\sigma_i$ is the corresponding cross section. The threshold energies for the BH process and meson production are $\epsilon_{\rm th, BH} = 2m_e$ and $\epsilon_{\rm th, meson} \approx 0.14$~GeV, respectively. The cooling timescales
$t_{{\rm cool}, i}(E_A)$ have a very similar expression to Eq.~(\ref{eq:trec}),
\begin{align}
t^{-1}_{{\rm cool}, i}(E_A) =& {c\over 2\Gamma_A^2} \int_{\epsilon_{{\rm th}, i}}^{2\Gamma_A E_\gamma^{\rm max}} d\epsilon_\gamma \epsilon_\gamma \sigma_i(\epsilon_\gamma)\kappa_i(\epsilon_\gamma) \nonumber \\ &\times \int^{E_\gamma^{\rm max}}_{\max[E_\gamma^{\rm min},  \epsilon_\gamma/(2\Gamma_A)]} dE_\gamma E_\gamma^{-2} n(E_\gamma),
\label{eq:tloss}
\end{align}
where $\kappa_i$ is the inelasticity (i.e., the energy loss fraction).

The photodisintegration is dominated by the excitation of the Giant Dipole Resonance (GDR). In the rest frame of the nuclei, the resonance energy $\epsilon_{\rm GDR} \approx 42.65A^{-0.21}$~MeV.
The corresponding GDR cross section is approximately  $\sigma_{\rm GDR} \approx 4.3 \times 10^{-28} A^{1.35}~{\rm cm}^2$, with an effective width $\Delta \epsilon_{\rm GDR} \approx 21.05A^{-0.35}$~MeV \cite{Zhang:2024sjp}. Assuming that only one nucleon is emitted, the inelasticity $\kappa_{\rm photodis} \approx 1/A$. With the resonance approximation, the timescales for photodisintegration given by Eqs.~\eqref{eq:trec} and \eqref{eq:tloss} can be simplified as \cite{Waxman:1997ti}
\begin{align}
& t_{\rm reac, photodis}^{-1}(E_A) = A t_{\rm cool, photodis}^{-1}(E_A) \nonumber  \nonumber \\
&\approx {U_\gamma c \sigma_{\rm GDR} \over \ln(E_{\gamma}^{\rm max}/E_{\gamma b}) E_{\gamma b}}
{\Delta \epsilon_{\rm GDR} \over \epsilon_{\rm GDR}} \min(\chi^{\alpha-1},\chi^{\beta-1}) \nonumber \\
& \approx 10^{-3}~{\rm s^{-1}}~ A^{1.21} (100~{\rm keV}/E^{\rm ob}_{\gamma b}) \nonumber \\ & \times L_{\gamma, \rm iso, 50} \Gamma_2^{-1} R_{14}^{-2}
\min(\chi^{\alpha-1},\chi^{\beta-1}),
\label{eq:t_photodis}
\end{align}
where $U_\gamma = L_{\gamma, \rm iso}/(4\pi\Gamma^2  R^2 c)$ is the photon energy density in the fluid rest frame, $\chi \equiv 2\Gamma_A E_{\gamma b}/\epsilon_{\rm GDR}$, and  $\ln(E_\gamma^{\rm max}/E_{\gamma b})\sim 10$. From Eq.~\eqref{eq:t_photodis}, we see that $t_{\rm reac, photodis}^{-1}$ and $t_{\rm cool, photodis}^{-1}$ grow linearly with $E_A\propto\chi$ ($\beta=2$) for $\chi < 1$, and decrease as $E_A^{-1/2}$ ($\alpha=0.5$) for $\chi > 1$. Note that Eq.~\eqref{eq:t_photodis} is only valid when the resonance is reached. 
If $E_A$ is too low, i.e., $(E_A/m_A) E_\gamma^{\rm max} < \epsilon_{\rm GDR}-0.5\Delta \epsilon_{\rm GDR}$, or too high, i.e., $(E_A/m_A) E_\gamma^{\rm min} > \epsilon_{\rm GDR}+0.5\Delta \epsilon_{\rm GDR}$, the resonance condition cannot be fulfilled, and the photodisintegration rates decrease rapidly.

For both the BH process and the photomeson production, we choose to solve Eqs.~\eqref{eq:trec} and \eqref{eq:tloss} numerically to obtain the relevant timescales. We take the inelasticity and cross section for the BH process from Ref.~\cite{Chodorowski:1992} [see its Eqs. (2.4), (2.5), (3.7), and (3.9)]. 

In the independent particle approximation, the photomeson production cross section is $\sigma_{\rm meson}(\epsilon_\gamma)\approx A \sigma_{p\gamma}(\epsilon_\gamma)$. The interaction among nucleons inside nuclei would mainly affect the cross section in two aspects. The nuclear medium tends to smear out the small resonances but has little impact on the overall cross section at low energies, whereas the total cross section is suppressed due to the nuclear shadowing at $\epsilon_\gamma \gtrsim 1$ GeV \cite{Weise1974,Arneodo:1992wf,Engel:1996yb}.
To properly account for these effects, we take the so-called empirical photomeson model from Ref.~\cite{Morejon:2019pfu} for the total photomeson cross section:
\begin{align}
\sigma_{\rm meson}(\epsilon_\gamma) \approx A^{\delta(\epsilon_\gamma)} \times \left\{ \begin{array}{cc} \sigma_{\rm univ}(\epsilon_\gamma), & \epsilon_\gamma \le 10~{\rm GeV},   \\ 
\sigma_{p\gamma}(\epsilon_\gamma), & \epsilon_\gamma > 10~{\rm GeV},
\end{array}
\right. 
\end{align}
where $\sigma_{\rm univ}$ is a universal cross section, the exponent $\delta(\epsilon_\gamma)$ of the nuclear mass number $A$ incorporates the nuclear medium effects with $\delta\approx 1$ at $\epsilon_\gamma \lesssim 1$ GeV and decreasing to $2/3$ in the high-energy limit, and $\sigma_{p\gamma}$ is the $p\gamma$ photomeson cross section. In our study, we use the values of $\sigma_{\rm univ}(\epsilon_\gamma)$ and $\delta(\epsilon_\gamma)$ from Fig.~3 of Ref.~\cite{Morejon:2019pfu} and $\sigma_{p\gamma}(\epsilon_\gamma)$ from SOPHIA \cite{Mucke:1999yb}.
Note that the empirical model of Ref.~\cite{Morejon:2019pfu} is constructed based on experimental data for nuclei with $A \le 56$. We assume that the model can be extended to describe the heavier $r$-process nuclei as well. Heavy nuclei fragment into lighter nuclei in the photomeson regime, so we take $\kappa_{\rm meson} = 1$ to estimate the cooling timescale $t_{\rm loss, meson}$ due to photomeson production.            

\begin{figure*}[htbp]  
\centering
\includegraphics[width=0.49\textwidth]{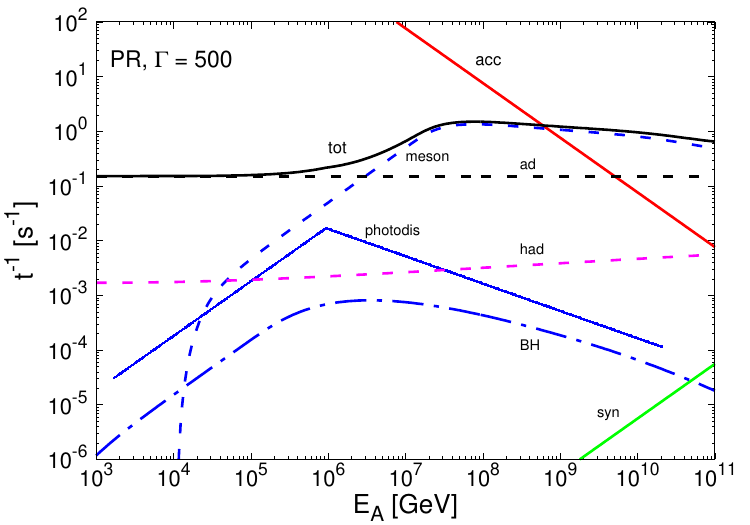} 
\includegraphics[width=0.49\textwidth]{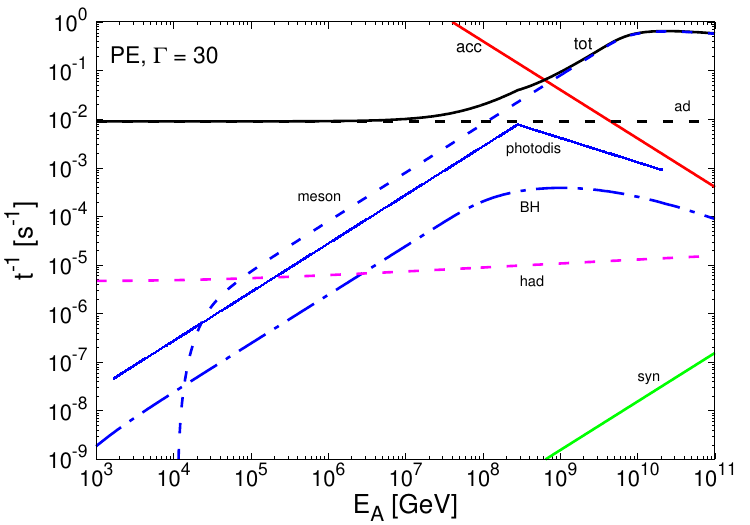} \\
\includegraphics[width=0.49\textwidth]{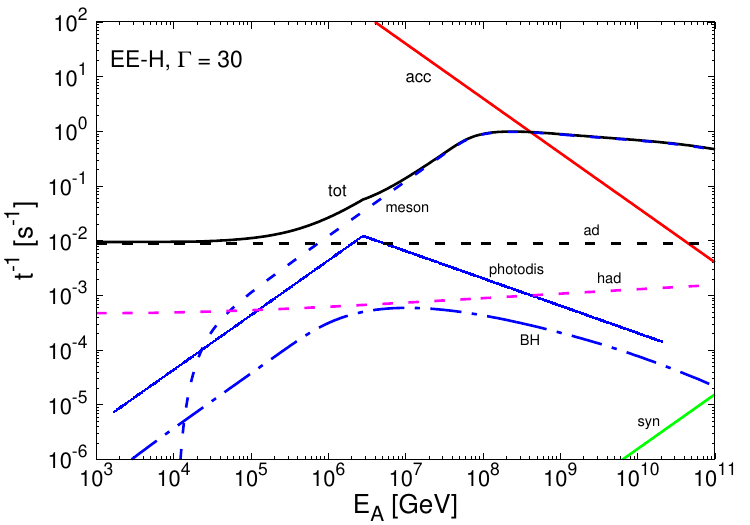} 
\includegraphics[width=0.49\textwidth]{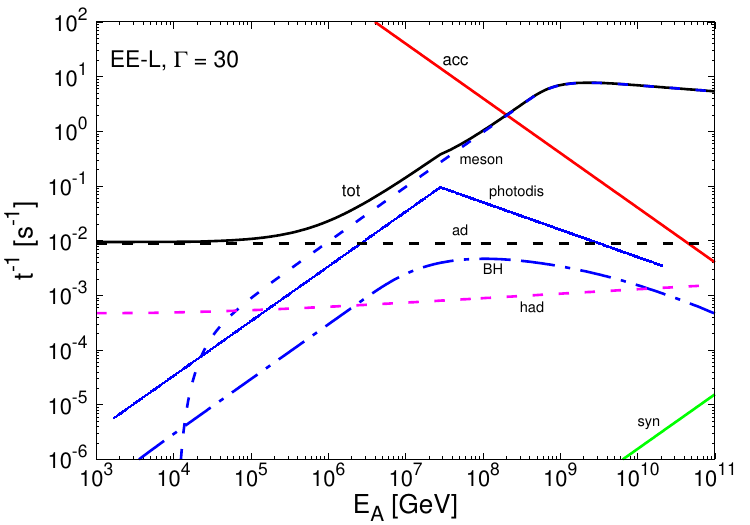}
\caption{The acceleration rate $t_{\rm acc}^{-1}$ (`acc') and the cooling rates for various processes are shown for the $r$-process nucleus, $^{130}_{~52}$Te, in the fluid frame. The total cooling rate
$t_{\rm cool}^{-1}$ (`tot') includes
$t_{\rm cool, meson}^{-1}$ (photomeson production, `meson'), $t_{\rm cool, photodis}^{-1}$ (photodisintegration, `photodis'), $t_{\rm cool, BH}^{-1}$ (Bethe-Heitler process, `BH'), $t_{\rm had}^{-1}$ (hadronic collision, `had'), $t_{\rm syn}^{-1}$ (synchrotron radiation, `syn'), and $t_{\rm ad}^{-1}$ (adiabatic expansion, `ad'). The upper left, upper right, lower left, and lower right panels correspond to the results for the PR, PE, EE-H, and EE-L cases, respectively.} 
\label{fig:timescales}  
\end{figure*}

Mainly for understanding the dependence on $\Gamma$ and $R$, we give the timescales for photomeson production in the $\Delta$-resonance limit, which are similar to those for photodisintegration in the GDR approximation [see Eq.~(\ref{eq:t_photodis})]:
\begin{align}
t^{-1}_{\rm cool, meson}(E_A) & \approx t^{-1}_{\rm reac, meson}(E_A) \nonumber \\ & \sim 2 \times 10^{-3}~{\rm s^{-1}}~A(100 {\rm~keV}/E_{\gamma b}^{\rm ob}) \nonumber \\ & \times L_{\gamma, \rm iso, 50} \Gamma_2^{-1} R_{14}^{-2} \min(\lambda^{-1/2}, \lambda), \label{eq:tpr_delta}
\end{align}
where $\lambda \equiv 2\Gamma_A E_{\gamma b}/\epsilon_\Delta$ with $\epsilon_\Delta \approx 0.3$ GeV, and where we have used $\alpha=0.5$ and $\beta=2$ explicitly. While the above approximation works for $\lambda<1$ and well above the meson production threshold, it breaks down for $\lambda>1$, in which regime $t^{-1}_{\rm cool, meson}$ only slowly decreases with $E_A$ due to channels other than the $\Delta$-resonance (see Fig.~\ref{fig:timescales}).

\section{The maximal energy and survival of accelerated nuclei}
\label{sec:Emax-survival}

In this section, we examine the conditions in various dissipation sites of sGRBs that are favorable for producing the most energetic CR events observed above $\sim 100$~EeV, which are hypothesized to consist primarily of $r$-process nuclei. 
Specifically, these regions
must be capable of accelerating heavy nuclei to ultra-high energies
while also ensuring
the survival of the accelerated nuclei during and after acceleration. 
In this work, we demonstrate that these two requirements can impose significant constraints
on key parameters of the dissipation sites, such as the bulk Lorentz factor ($\Gamma$), the dissipation radius ($R$), the magnetic field strength, and the acceleration efficiency.

\subsection{The maximal energy of accelerated nuclei}

The maximal energy of the accelerated nuclei is limited by the cooling and escape processes. For the cooling of heavy nuclei, we consider all 
three $A\gamma$ reaction channels discussed in Sec.~\ref{sec:timescale} [see Eq.~\eqref{eq:tloss}], hadronic collisions with protons, synchrotron radiation in the magnetic field, and adiabatic expansion. The total cooling rate for heavy nuclei is $t_{\rm cool}^{-1}=t_{\rm cool, BH}^{-1}+t_{\rm cool, photodis}^{-1}+t_{\rm cool, meson}^{-1}+t_{\rm had}^{-1}+t_{\rm syn}^{-1}+t_{\rm ad}^{-1}$.
The cooling timescales for the last three processes are: 
\begin{align}
& t_{\rm had}(E_A) \approx {1\over \sigma_{Ap}n_p c}, \\
& t_{\rm syn}(E_A) \approx {6\pi m_A^4 c^3\over Z^2 \sigma_{\rm Th} m_e^2 E_A B^2}, \label{eq:syn}\\
& t_{\rm ad}(E_A) \approx {R\over \Gamma c}. \label{eq:tad}
\end{align}
Here, $n_p \sim L_{\rm k, iso}/(4\pi \Gamma^2 R^2 m_p c)$ is the thermal proton number density inside the jets \cite{Kumar2014}, with $L_{\rm k, iso}$ representing the jet kinetic energy, $\sigma_{Ap}$ is the inelastic proton-nucleus cross section, and $\sigma_{\rm Th}\approx 6.65\times 10^{-25}~{\rm cm}^2$ is the Thomson cross section. We take 
$L_{ \rm k, iso} = 10 L_{\gamma, \rm iso}$.
In the relevant energy range, $\sigma_{Ap}(s_{Ap}) \approx A^{2/3}\sigma_{pp}(s_{pp}\approx s_{Ap}/A)$, where $s_{Ap}$ and $s_{pp}$ are the center-of-mass energy squared for the $A$-$p$ and $p$-$p$ pair, respectively  \cite{Guzey:2005tk,Wibig:1998by}. For inelastic $pp$ collisions, we use $\sigma_{pp}(s) \approx [32.4-1.2\ln(s)+0.21\ln^2(s)]$~mb, with $s$ in GeV$^2$ \cite{Wibig:1998by}.

As the escape timescale is typically longer than the dynamical timescale $t_{\rm dyn}\approx t_{\rm ad}$, we do not need to consider the escape process explicitly when calculating the maximal energy of accelerated nuclei \cite{Baerwald:2013pu,Biehl:2017zlw}. Consequently, the maximum energy, 
$E_{A, \rm max}$, can be determined by equating the acceleration and cooling timescales, i.e., $t_{\rm acc}(E_{A, \rm max})=t_{\rm cool}(E_{A, \rm max})$. For relativistic jets, the acceleration timescale for nuclei in the fluid frame can be expressed as
\begin{align}
t_{\rm acc}(E_A) = { \kappa_{\rm acc} E_A\over Ze c B}, \label{eq:tacc}
\end{align}
where the acceleration efficiency $\kappa_{\rm acc}$ incorporates the complex and uncertain physical processes associated with particle acceleration, $Z$ is the nuclear charge number, and $B = (8\pi U_B)^{1/2} = (8\pi U_\gamma \xi_{B/e})^{1/2}$ is the magnetic field strength in the dissipating fluid, with $\xi_{B/e}\equiv U_B/U_\gamma$ being the ratio of energy dissipated into magnetic field to that into electrons. In the Bohm limit, $\kappa_{\rm acc}$ is close to 1, corresponding to the maximum acceleration efficiency. In more realistic scenarios, such as those involving diffusive shock acceleration, $\kappa_{\rm acc}$  falls within the range of 10--100 (see, e.g., \cite{Globus:2014fka}).

\begin{figure*}[htbp]  
\centering
\includegraphics[width=0.49\textwidth]{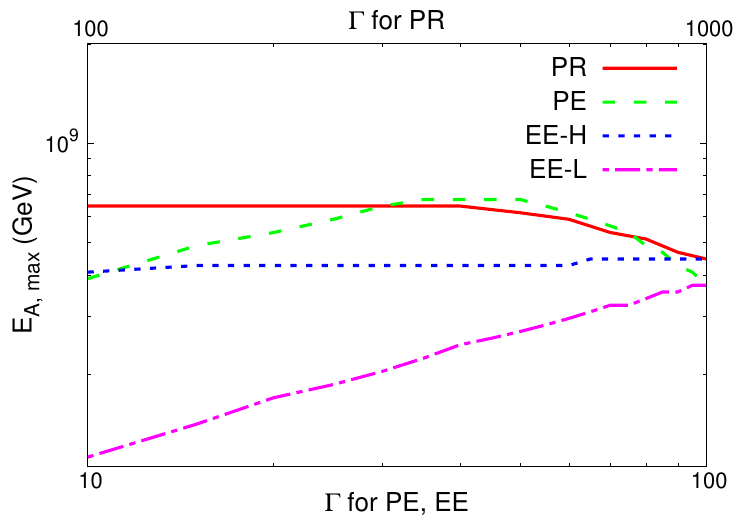} 
\includegraphics[width=0.48\textwidth]{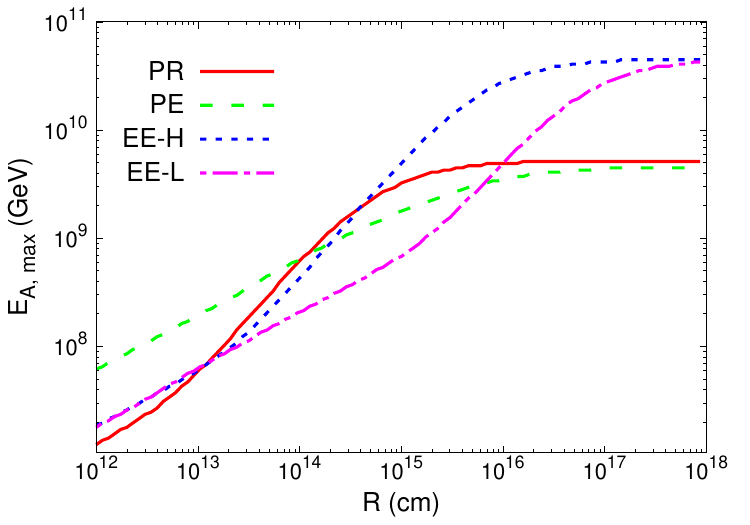}
\caption{$E_{A, \rm max}$ as functions of $\Gamma$ and $R$ for different emission cases. In the left panel, $R$ is set to $10^{14}$~cm. In the right panel, $\Gamma$ is fixed at 500 for PR, and at 30 for PE, EE-H, and EE-L. The heavy nuclei are assumed to be $^{130}_{~52}$Te.} 
\label{fig:Epmax}  
\end{figure*} 

Fig.~\ref{fig:timescales} presents the relevant timescales for $r$-process nuclei in the comoving frame.
Throughout this work, we consider tellurium ($^{130}_{~52}$Te)
as a representative $r$-process nucleus. For demonstration, we use $\Gamma = 500$ for PR and $\Gamma = 30$ for PE, EE-H, and EE-L. For all emission phases, we set $R = 10^{14}$~cm, $\kappa_{\rm acc} = 10$, and $\xi_{B/e}=1$. The results indicate that photomeson production, photodisintegration, and adiabatic expansion are the dominant cooling processes, while other processes such as hadronic collisions, the BH process, and synchrotron cooling are comparatively negligible. Note that the results of $t_{\rm cool, meson}^{-1}$ shown in Fig.~\ref{fig:timescales}
agree well with those derived in the $\Delta$-resonance limit for $\lambda < 1$ and well above the meson production threshold [see Eq.~\eqref{eq:tpr_delta}]. As mentioned above, at energies above $E_{A}^{\rm pk} \approx 0.5 m_A \epsilon_\Delta/E_{\gamma b}$ (i.e., $\lambda > 1$), channels other than the $\Delta$-resonance are significant, and the exact calculations show that the photomeson production rates decrease more slowly than $E_A^{-1/2}$ as given in Eq.~\eqref{eq:tpr_delta}. A better approximation is to replace the last factor in Eq.~(\ref{eq:tpr_delta}) with $\min(1, \lambda)$.

For the selected parameters in Fig.~\ref{fig:timescales}, the intersection between $t^{-1}_{\rm cool}$ (black solid line) and $t^{-1}_{\rm acc}$ (red solid line) gives $E_{A, {\rm max}}^{\rm ob} = \Gamma E_{A,{\rm max}} \approx 3.1 \times 10^{11}$, $2 \times 10^{10}$, $10^{10}$, and $6\times 10^9$~GeV for PR, PE, EE-H, and EE-L, respectively. The variations of $E_{A, {\rm max}}$ with $\Gamma$ and $R$ are shown in 
Fig.~\ref{fig:Epmax}. In the left panel, $R$ is set to $10^{14}$~cm, and in the right panel, $\Gamma=500$ for PR and $\Gamma=30$ for PE, EE-H, and EE-L. The dependence of $E_{A, {\rm max}}$ on $\Gamma$ and $R$ can be understood as follows. When meson production dominates the cooling of heavy nuclei for small $\Gamma$ and $R$, the equality $t_{\rm acc}^{-1} = t^{-1}_{\rm cool, meson}$ gives $E_{A, \rm max} \propto L_{\gamma, \rm iso}^{-1/4} \xi_{B/e}^{1/4}\kappa_{\rm acc}^{-1/2} \Gamma^{1/2} R^{1/2}$ (EE-L and PE with smaller $E_{\gamma b}^{\rm ob}$) or $E_{A, \rm max} \propto L_{\gamma, \rm iso}^{-1/2} \xi_{B/e}^{1/2}\kappa^{-1}_{\rm acc} R$ (PR and EE-H with larger $E_{\gamma b}^{\rm ob}$) as shown in both panels of Fig.~\ref{fig:Epmax}. When $\Gamma$ and $R$ are large, adiabatic cooling takes over (i.e., $t_{\rm ad}^{-1}>t_{\rm cool,meson}^{-1}$). In this regime, $E_{A, \rm max} \propto L_{\gamma, \rm iso}^{1/2} \xi_{B/e}^{1/2}\kappa_{\rm acc}^{-1}\Gamma^{-2}$ (see PR and PE with larger $\Gamma$ in the left panel of Fig.~\ref{fig:Epmax}, and all four cases with larger $R$ in the right panel). With the scaling relations given above, the trends of $E_{A, {\rm max}}$ with $L_{\gamma, \rm iso}$ can also be obtained in a straightforward manner (not shown).  

\subsection{The survival of accelerated nuclei} 

To significantly contribute to the observed UHECR events, particularly the  `Amaterasu' event, the accelerated heavy nuclei must not only survive interactions with the intense photon background during and after the acceleration process, but also escape from the acceleration region when the appropriate energy is reached. The latter requirement may impose separate constraints on the parameters of sGRB jets. However, due to the complexity and uncertainties involved in modeling particle escape processes \cite{Baerwald:2013pu,Biehl:2017zlw,Zhang:2017moz,Globus:2023bhr}, we do not explore this aspect in this work. Here we simply require that the accelerated nuclei survive the $A\gamma$ reactions within the typical dynamical timescale $t_{\rm dyn}\approx R/(\Gamma c)$, which represents the minimum time needed for escape \cite{Zhang:2017moz}. As such, our constraints are based on the requirements for efficient acceleration and survival of heavy nuclei, and should be considered conservative.

In the dissipation regions of sGRBs, where $E_\gamma \gg 1$~eV, the photomeson production is the dominant process that breaks UHE heavy nuclei into smaller fragments. We introduce the ``optical depth'' for photomeson production as \cite{Zhang:2017moz}   
\begin{align}
\tau_{A\gamma}(E_A) \approx {t_{\rm dyn} \over t_{\rm reac, meson}(E_{A})} \approx {R \over \Gamma c t_{\rm reac, meson}(E_{A})}. \label{eq:tau_req}
\end{align}
The survival of heavy nuclei corresponds to a small optical depth $\tau_{A\gamma} \lesssim 1$.

\begin{figure*}[htbp]  
\centering
\includegraphics[width=0.49\textwidth]{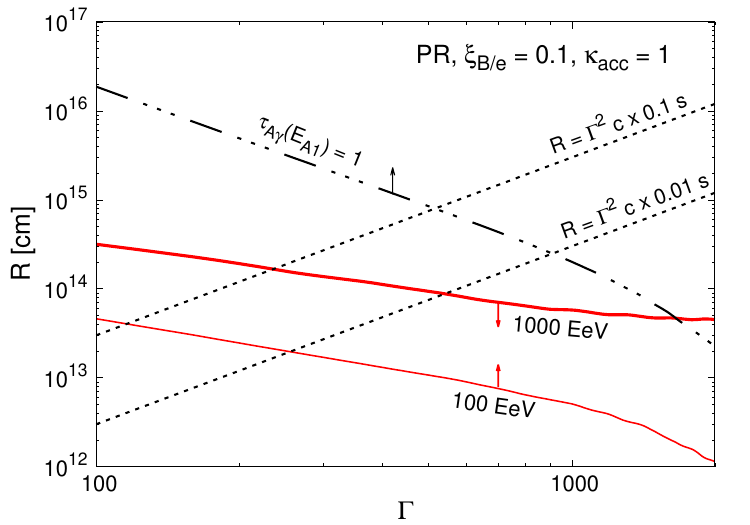} 
\includegraphics[width=0.49\textwidth]{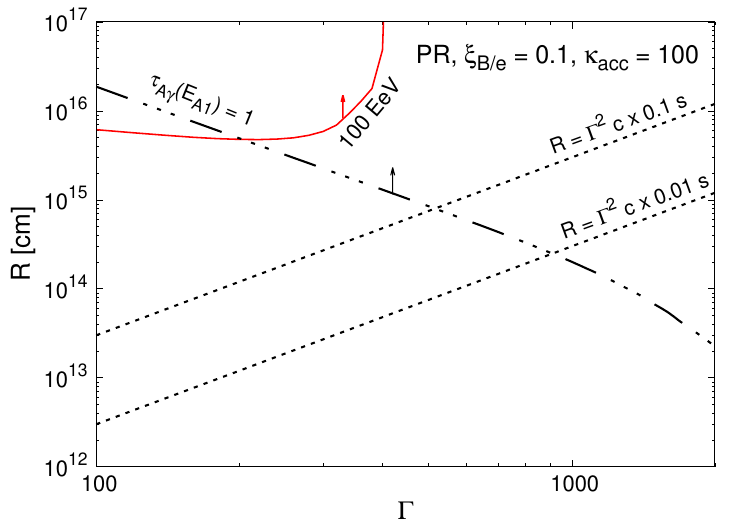} \\
\includegraphics[width=0.49\textwidth]{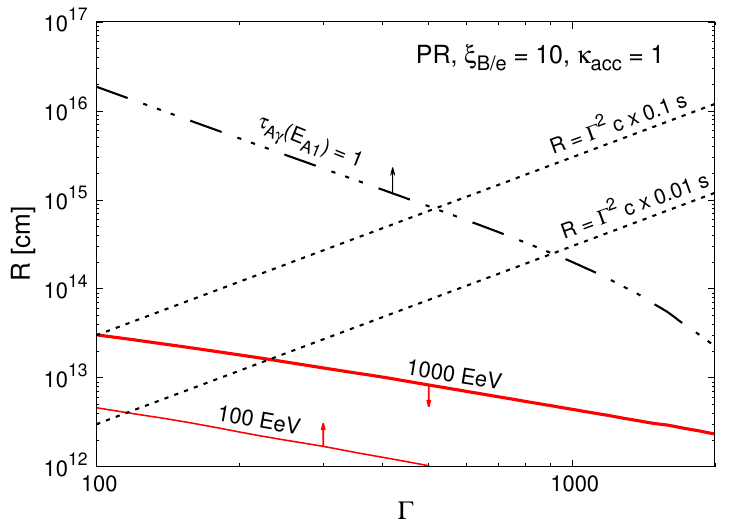} 
\includegraphics[width=0.49\textwidth]{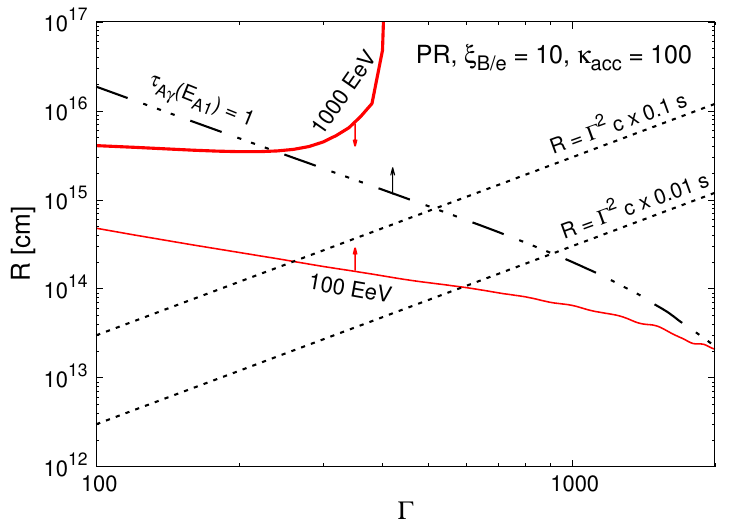}
\caption{Contours in the $\Gamma$-$R$ plane corresponding to $\tau_{A\gamma}(E_{A1})=1$, and $E_{A, \rm max}^{\rm ob}=100$ and 1000 EeV for the PR phase. Four different combinations of $\xi_{B/e}$ and $\kappa_{\rm acc}$ are considered: $(\xi_{B/e}, \kappa_{\rm acc})= (0.1, 1)$ (upper left), (0.1, 100) (upper right), (10, 1) (lower left), and (10, 100) (lower right). The heavy nuclei are assumed to be $^{130}_{~52}$Te. Additionally, the expected radii of internal shocks, approximated as $R \approx \Gamma^2 c t_v$, are indicated (black dashed lines), where the variability timescale $t_v$ is set to 0.01 and 0.1 s, representing the minimum and maximum values, respectively. The arrows point to the parameter regions corresponding to the requirements in Eq.~\eqref{eq:req-1}. See text for details.} 
\label{fig:LFR_range}  
\end{figure*} 

\subsection{Constraints on jet parameters}

Assuming that $E_{A, \rm max}$ is
proportional to the nuclear mass number $A$, and considering various compositions of conventional heavy nuclei, fits to the Auger data suggest $E_{A, \rm max}/A \approx 1$--2 EeV \cite{Auger:2016use,PierreAuger:2023htc}, corresponding to a maximum energy around 100 EeV for iron nuclei.
Ref.~\cite{Zhang:2024sjp} found that including contributions from $r$-process nuclei (with $A\sim 100$--200) could lead to a better fit to the TA data, and
the maximal energy of these $r$-process nuclei escaping from the sources typically reaches several hundred EeV under various assumptions. For sGRBs to be considered as potential sources of the most energetic CR events around a few hundred EeV, we impose the following requirements:
\begin{align}
E_{A1}^{\rm ob} < E_{A, \rm max}^{\rm ob} < E_{A2}^{\rm ob}~~{\rm and}~~ \tau_{A\gamma}(E_{A1}) < 1, \label{eq:req-1}  
\end{align}
where $E_{A1}^{\rm ob}=\Gamma E_{A1}=100$~EeV and $E_{A2}^{\rm ob}=\Gamma E_{A2}=1000$~EeV.
The above requirements ensure that heavy nuclei accelerated to the maximal energy of 100--1000 EeV, as inferred from Ref.~\cite{Zhang:2024sjp}, survive with sufficiently high probability to contribute to the observed UHECRs.

\begin{figure*}[htbp]  
\centering
\includegraphics[width=0.49\textwidth]{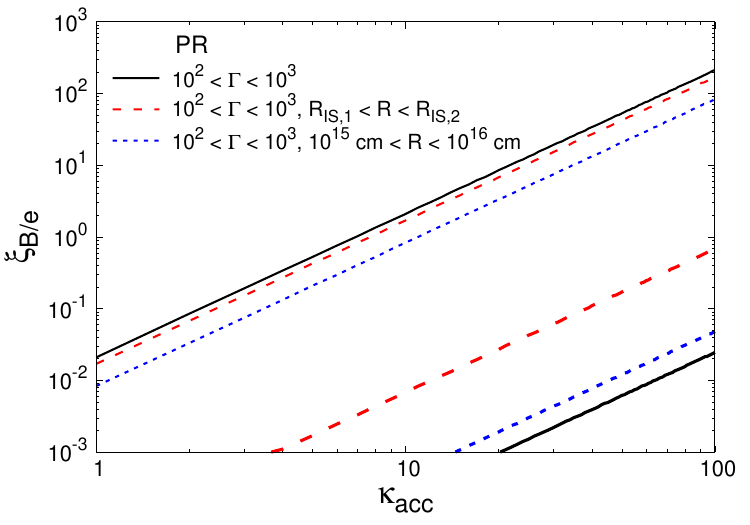} 
\includegraphics[width=0.49\textwidth]{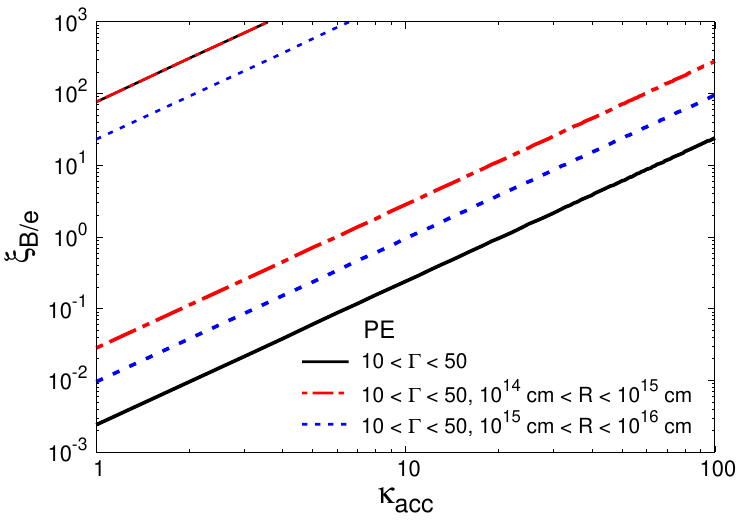} \\
\includegraphics[width=0.49\textwidth]{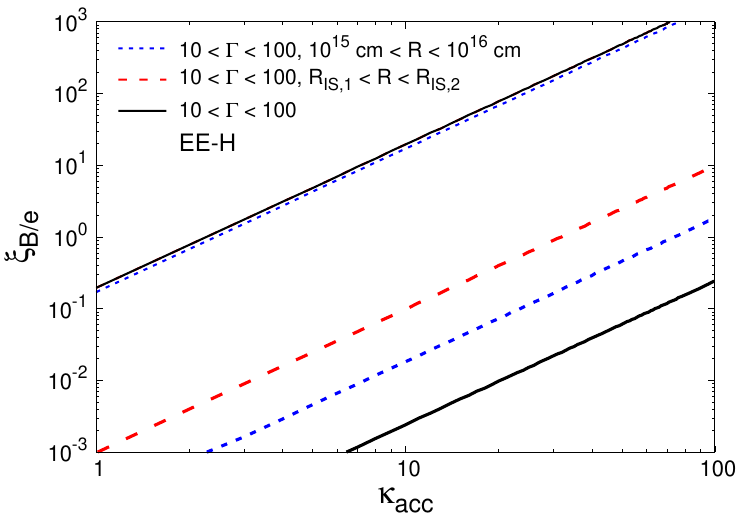} 
\includegraphics[width=0.49\textwidth]{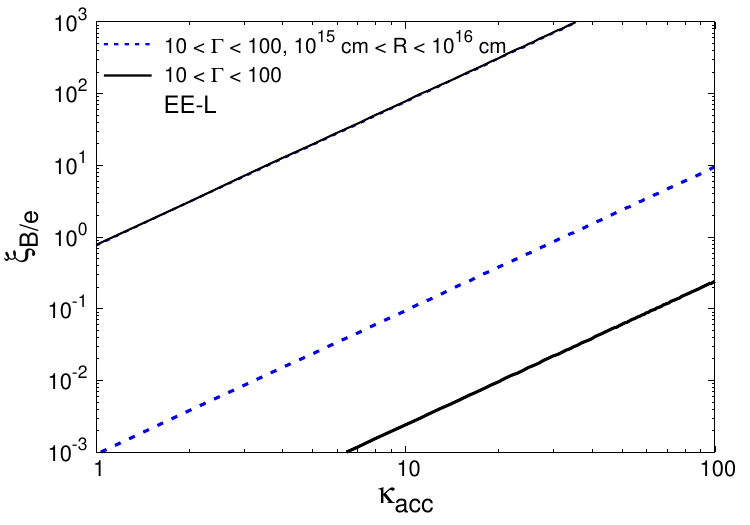}
\caption{Allowed regions in the $\kappa_{\rm acc}$--$\xi_{B/e}$ plane for sGRBs to contribute significantly to the most energetic CR events when different additional constraints on $\Gamma$ and $R$ are considered for the PR, PE, EE-H, and EE-L cases. The thin and thick lines represent the upper and lower bounds on $\xi_{B/e}$ as functions of $\kappa_{\rm acc}$. The heavy nuclei are assumed to be $^{130}_{~52}$Te. See text for details.}  
\label{fig:kap-Xibe}  
\end{figure*}

With the photon luminosity fixed for different dissipation sites (see Table~\ref{tab:pars}), the optical depth $\tau_{A\gamma}$ depends on $\Gamma$ and $R$, while $E^{\rm ob}_{A, \rm max}$ varies with $\Gamma$, $R$, $\kappa_{\rm acc}$, and $\xi_{B/e}$. These parameters can be constrained by Eq.~\eqref{eq:req-1}. Taking the PR phase for illustration, we show the contours in the $\Gamma$--$R$ plane for $\tau_{A\gamma}(E_{A1})=1$, $E_{A, \rm max}^{\rm ob}=E_{A1}^{\rm ob}$, and $E_{A, \rm max}^{\rm ob}=E_{A2}^{\rm ob}$, for four representative combinations of
$(\xi_{B/e}, \kappa_{\rm acc})$: (0.1, 1) (upper left panel of Fig.~\ref{fig:LFR_range}), (0.1, 100) (upper right), (10, 1) (lower left), and (10, 100) (lower right). 

Over the range of $\Gamma=100$--1000 considered for the PR phase (see Table~\ref{tab:pars}), $\lambda=2E_A^{\rm ob} E_{\gamma b}^{\rm ob}/(\Gamma^2 m_A\epsilon_\Delta)>1$ and
$\tau_{A\gamma} \propto L_{\gamma,\rm iso}R^{-1} \Gamma^{-2}$ is approximately independent of $E_A$ [see Eq.~\eqref{eq:tpr_delta} for the rate of cooling by meson production and related discussion of Fig.~\ref{fig:timescales}]. So the contour for $\tau_{A\gamma}(E_{A1})=1$ follows $R\propto\Gamma^{-2}$. Below this contour, the most energetic heavy nuclei are unlikely to survive. This result alone can already be used to impose significant constraints on those sGRBs that contribute to the most energetic CR events. Specifically, in the internal shock (IS) model, the dissipation radius is approximately $R \approx \Gamma^2ct_v$, where $t_v$ is the variability timescale. The typical IS radii for the PR phase (black dashed lines) are shown in Fig.~\ref{fig:LFR_range}, with $t_v = 0.01$ and 0.1 s. As can be seen, for the IS model, $\Gamma$ must exceed $\sim 500$ to prevent the destruction of UHE heavy nuclei. For the internal collision-induced magnetic reconnection and turbulence (ICMART) model with larger emission radii \cite{Zhang:2011}, the constraint on $\Gamma$ can be relaxed. For instance, with $R = 2\times 10^{15}$~cm, $\Gamma$ is constrained to be $\gtrsim 400$.

When cooling by meson production dominates (for small $\Gamma$ or $R$), $E_{A, \rm max}^{\rm ob} = \Gamma E_{A, \rm max} \propto \xi_{B/e}^{1/2}\kappa_{\rm acc}^{-1}\Gamma R$ ($\lambda>1$), and the contours of $E_{A, \rm max}^{\rm ob}=100$ and 1000 EeV in the $\Gamma$--$R$ plane approximately follow $R \propto \Gamma^{-1}$ (the left panels of Fig.~\ref{fig:LFR_range}, and $E_{A, \rm max}^{\rm ob}=100$ EeV in the lower right panel). When adiabatic cooling dominates (for sufficiently large $\Gamma$ or $R$), $E_{A, \rm max}^{\rm ob} \propto \xi_{B/e}^{1/2}\kappa_{\rm acc}^{-1}\Gamma^{-1}$, and the contours become vertical (corresponding to constant $\Gamma$, see $E_{A, \rm max}^{\rm ob}=100$ EeV in the upper right panel and $E_{A, \rm max}^{\rm ob}=1000$ EeV in the lower right panel). The vertical contours are connected to nearly horizontal ones (corresponding to constant $R$), for which cooling by meson production is comparable to adiabatic cooling.

To satisfy the requirements in Eq.~\eqref{eq:req-1}, the allowed $R$ and $\Gamma$ must lie above the contour of $\tau_{A\gamma}(E_{A1})=1$ and fall within the bounds set by $E_{A, \rm max}^{\rm ob}=E_{A1}^{\rm ob}$ and $E_{A, \rm max}^{\rm ob}=E_{A2}^{\rm ob}$. As shown in the left panels of Fig.~\ref{fig:LFR_range}, with $\kappa_{\rm acc}=1$ and $\xi_{B/e}=0.1$ or 10, there is no parameter space in the $\Gamma$--$R$ plane that meets the requirements for the PR case with $\Gamma=100$--1000. With $\xi_{B/e}=10$ and $\kappa_{\rm acc}=100$ (lower right panel), a considerable region in the $\Gamma$-$R$ plane fulfills the conditions for producing and preserving UHE heavy nuclei. This region is also consistent with the typical parameters for the IS and ICMART models. Although another allowed region exists for $\xi_{B/e}=0.1$ and $\kappa_{\rm acc}=100$ (upper right panel), the corresponding radii are significantly larger than typically expected for the IS and ICMART models.

The impact of $\xi_{B/e}$ and $\kappa_{\rm acc}$ arises from their influence on $E_{A, \rm max}^{\rm ob}$.
As $E_{A, \rm max}^{\rm ob}\propto\xi_{B/e}^{1/2}\kappa_{\rm acc}^{-1}\Gamma R$ for cooling by meson production with $\lambda>1$ and $E_{A, \rm max}^{\rm ob}\propto\xi_{B/e}^{1/2}\kappa_{\rm acc}^{-1}\Gamma^{-1}$ for adiabatic cooling, changing $\xi_{B/e}^{1/2}\kappa_{\rm acc}^{-1}$ shifts the contours of $E_{A, \rm max}^{\rm ob}$ in the $\Gamma$-$R$ plane. Because 
$\tau_{A\gamma}\propto R^{-1}\Gamma^{-2}$, depending on the assumed values of $\kappa_{\rm acc}$ and $\xi_{B/e}$, there may or may not be regions in the $\Gamma$-$R$ plane that satisfy Eq.~\eqref{eq:req-1}, as shown in Fig.~\ref{fig:LFR_range}. By scanning over $\Gamma$ and $R$, the projected region in the $\kappa_{\rm acc}$--$\xi_{B/e}$ plane that satisfies Eq.~\eqref{eq:req-1} for the PR case is shown in the upper left panel of Fig.~\ref{fig:kap-Xibe}, where thin and thick lines denote the upper and lower bounds, respectively, on $\xi_{B/e}$ as functions of $\kappa_{\rm acc}$. In addition to the restriction of $\Gamma=100$--1000 (Table~\ref{tab:pars}), projections are also made by incorporating $R_{\rm IS, 1}<R<R_{\rm IS, 2}$ for the IS model (see also Fig.~\ref{fig:LFR_range}) and $10^{15}~{\rm cm}<R<10^{16}~{\rm cm}$ for the ICMART model. As expected, applying these additional constraints on $R$ reduces the viable space in the $\kappa_{\rm acc}$--$\xi_{B/e}$ plane. Nevertheless, typical values of $\xi_{B/e}\sim0.1$--10 and $\kappa\sim10$--100 are still allowed, supporting the possibility that UHE heavy nuclei can be produced during the PR phase of sGRBs.

Similar discussion to that presented above for the PR phase can be extended to other emission phases. The constraints on $\Gamma$ and $R$ for the PE, EE-H, and EE-L cases are shown in Fig.~\ref{fig:LFR_range_all} (see Appendix~\ref{app:constraints} for details). We emphasize that
the radii required to preserve UHE heavy nuclei in the EE phase are larger than typically assumed in the literature. Specifically, $R$ must exceed $\sim 10^{15}$~cm for the EE-H and EE-L cases, and $\sim 10^{14}$~cm for the PE case. The projected allowed regions in the $\kappa_{\rm acc}$--$\xi_{B/e}$ plane for the PE, EE-H, and EE-L cases are shown in Fig.~\ref{fig:kap-Xibe} by incorporating various constraints on $R$. Similar to the PR case, relatively broad ranges of $\kappa_{\rm acc}$ and $\xi_{B/e}$ are allowed for these cases as well. However, for a given $\kappa_{\rm acc}$, the allowed $\xi_{B/e}$ for the PE phase is substantially higher than that for the PR and EE phases. 

Reference~\cite{Gompertz:2015} investigated the broad-band spectrum during the PE phase, considering late-time energy injection into the forward shock from a magnetar, and determined that $\epsilon_{B}$ ranges from $10^{-7}$ to 1, while $\epsilon_e$ typically lies within 0.1--1. These inferred values are consistent with afterglow observations, indicating much lower values of $\epsilon_B$ compared to $\epsilon_e$ \cite{Kumar2014}. Assuming $\xi_{B/e} \lesssim 0.1$ for most sGRBs during the PE phase, our analysis suggests that this phase is unlikely to contribute UHE heavy nuclei, unless $\kappa_{\rm acc} \sim 1$ is realized by e.g., rapid particle acceleration in highly turbulent environments.

To briefly summarize this section, we find that the PR phase, with $\Gamma \gtrsim 400$-500 and typical values of $\kappa_{\rm acc}$ and $\xi_{B/e}$, is well suited for contributing UHE $r$-process nuclei. In contrast, the EE phase requires dissipation radii larger than typically adopted in the literature to ensure the survival of UHE heavy nuclei, while the PE phase demands highly efficient particle acceleration to reach the required energies. These limitations suggest that the EE and PE phases are less favored as dominant sources of UHE $r$-process nuclei.

\begin{table*}[htbp] 
\centering 
\renewcommand{\arraystretch}{1.5}
\caption{Parameters for computing the HE neutrino fluences from different emission phases of sGRBs. For each emission phase, three different sets of $(\Gamma, R)$ are considered, chosen such that the optical depth $\tau_{A\gamma}(E_{A1})=1$. We have considered two extreme compositions, for which the accelerated nuclei consist of either pure protons or solely $r$-process nuclei ($^{130}_{~52}$Te). The corresponding isotropic spectra follow power-law distributions, with the normalization constant $C_{p0}$ ($C_{A0}$) and the maximum energy $E_{p, \rm max}$ ($E_{A, \rm max}$) [Eqs.~\eqref{eq:pspec} and \eqref{eq:Aspec}]. The values of $E^{\rm ob}_{A,\rm max}=\Gamma E_{A,\rm max}$ and $E^{\rm ob}_{p,\rm max}=\Gamma E_{p,\rm max}$ are shown in the last column for corresponding sets of $(\Gamma, R)$.    
\label{tab:pars-nu}}
\begin{ruledtabular}
\begin{tabular}{cccccc}    
Parameters &
$C_{p0} \approx C_{A0}$ (GeV) & $\kappa_{\rm acc}$ & $\xi_{B/e}$  & $(\Gamma, R/10^{14}~{\rm cm})$ & $(E^{\rm ob}_{A,\rm max}, E^{\rm ob}_{p,\rm max})/{\rm EeV}$ \\    
\hline
PR & $4.5\times 10^{54}\Gamma^{-1}$ & 10 & 0.1 & (300, 22.6), (500, 8.50), (800, 3.30) & $(771,12.2),(446,9.31),(259,8.19)$  \\ 
EE-H & $ 10^{54}\Gamma^{-1}$ & 10 & 0.1 & (20, 184), (30, 85.2), (50, 32.4) & $(356,5.38),(222,3.69),(135,3.23)$ \\
EE-L & $10^{54}\Gamma^{-1}$ & 10 & 0.1 & (20, 1852), (30, 851), (50, 280) & $(340,7.43),(222,6.41),(129,4.67)$ \\ 
PE & $10^{53}\Gamma^{-1}$ & 10 & 10 & (20, 134), (30, 30.0), (50, 3.75) & $(296,11.2),(169,7.71),(117,4.67)$ \\
\end{tabular}
\end{ruledtabular}
\end{table*}

\begin{figure*}[htbp]  
\centering
\includegraphics[width=0.49\textwidth]{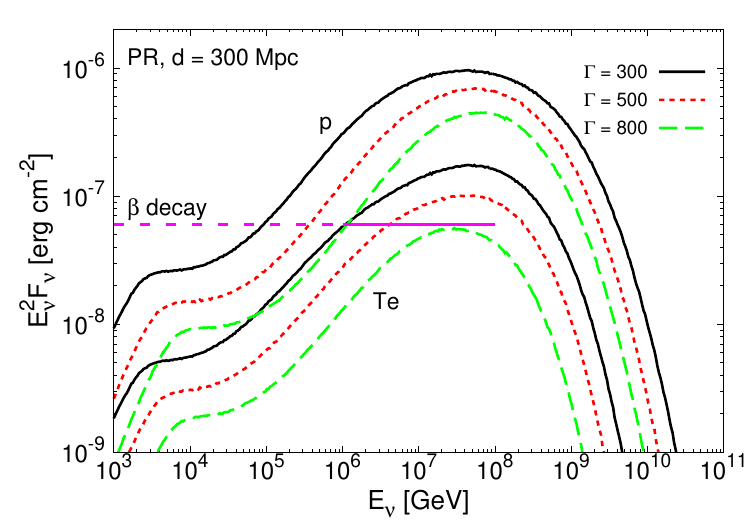} 
\includegraphics[width=0.49\textwidth]{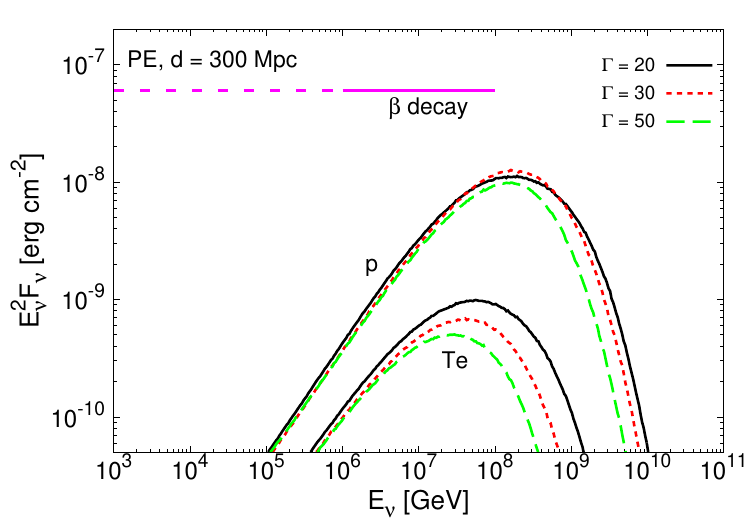} \\
\includegraphics[width=0.49\textwidth]{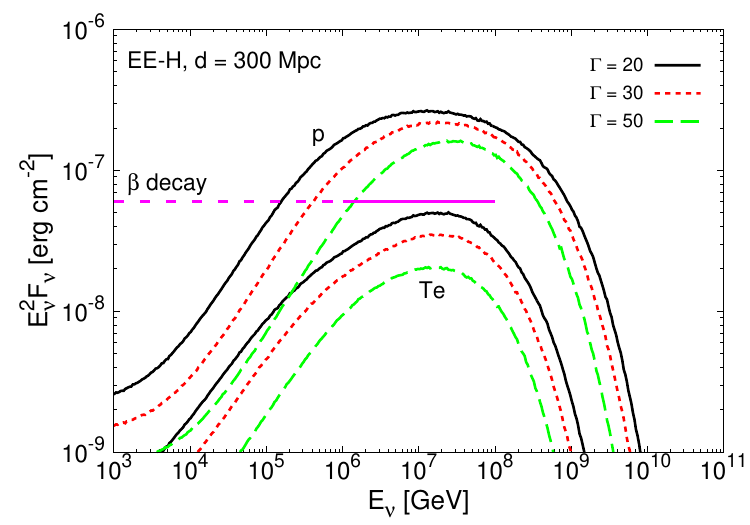} 
\includegraphics[width=0.49\textwidth]{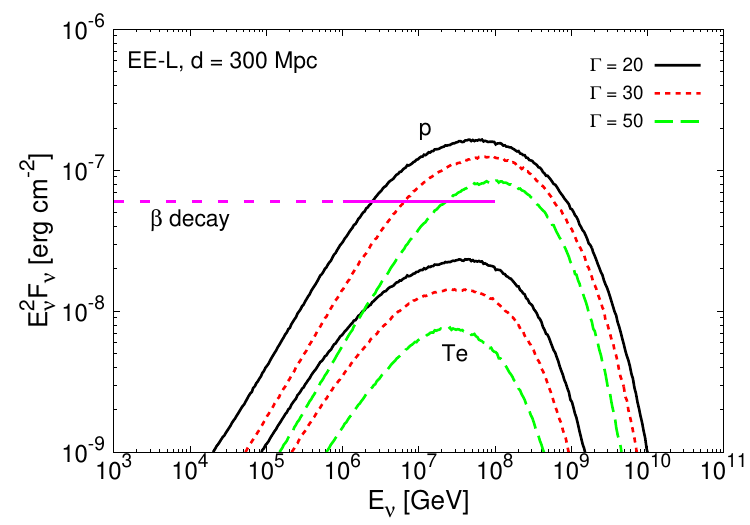}
\caption{All-flavor neutrino fluences from an sGRB located at $d = 300$ Mpc during different emission phases: PR (upper left), PE (upper right), EE-H (lower left), and EE-L (lower right). The curves represent neutrino fluences from the photomeson and hadronic interactions of protons and $^{130}_{~52}$Te using different sets of $(\Gamma, R)$. For each emission phase, three representative values of $\Gamma$ are considered, with the corresponding radius $R$ chosen to give $\tau(E_{A1})=1$ (see Table~\ref{tab:pars-nu}).
The magenta lines indicate neutrino fluences from $\beta$ decay of radioactive $r$-process nuclei, normalized using the observed UHECR data above 100 EeV and assumed to be the same across all emission phases. The energy of $\beta$ decay neutrinos typically falls within 1--100 PeV (solid segments), but the spectrum may extend to lower energies (dashed segment) if lower-energy $r$-process nuclei can also escape from the source. See text for details.}   
\label{fig:nuflx}  
\end{figure*}

\section{HE neutrinos from sGRBs}\label{sec:neutrinos}

In this section, we study the expected HE neutrino signals from sGRBs across different emission phases, using parameter sets that support potential contribution to UHE heavy nuclei. We further discuss these signals as probes of the origin of UHECRs, particularly the most energetic events.        

\subsection{Limiting compositions and spectra of accelerated nuclei} 

The HE neutrino fluxes from sGRBs depend on both the fluence and the composition of the accelerated nuclei. We normalize the spectrum of these nuclei using the observed gamma-ray luminosities. The total isotropic energy in accelerated nuclei per sGRB can be estimated as $\mathcal{E}_{\rm cr, iso} \sim  \mathcal{E}_{\gamma, \rm iso} \eta_{\rm cr}$, where $\mathcal{E}_{\gamma, \rm iso}$ is the total isotropic photon energy (see Table~\ref{tab:pars}),
and $\eta_{\rm cr}$ represents the ratio of energy carried by accelerated nuclei relative to photons.\footnote{As noted below Eq.~\eqref{eq:tad}, we assume a kinetic luminosity $L_{\rm k,iso}=10L_{\gamma, \rm iso}$ to estimate the hadronic timescale, implying that the thermal baryons in the jets carry ten times more energy than the emitted gamma-rays. Considering that the energy carried by accelerated (non-thermal) nuclei could be comparable to that of 
the thermal ones, we adopt $\eta_{\rm cr}=10$.} Assuming $\eta_{\rm cr}= 10$ for all emission phases, the total energies of accelerated nuclei are $\mathcal{E}_{\rm cr, iso} \sim 10^{53}$~erg for the PR case, $\sim 3\times 10^{52}$~erg for the EE-H and EE-L cases, and $\sim 3\times 10^{51}$~erg for the PE case. The actual composition of accelerated nuclei is uncertain, so we consider two extreme limits to compute the HE neutrino fluxes for all emission phases. Specifically, we assume that the accelerated nuclei are either pure protons or solely of the $r$-process origin. The pure proton limit represents the conventional composition, and the HE neutrino fluxes may decrease by a factor of a few if a more realistic composition is considered \cite{Biehl:2017zlw,Morejon:2019pfu,DeLia:2024kjv}. As previously mentioned, we take $^{130}_{~52}$Te as a representative $r$-process nucleus. The actual HE neutrino fluxes from a source of UHE $r$-process nuclei should lie between those calculated for the above two limiting compositions.

The isotropic energy spectra of HE protons and $r$-process nuclei (in units of GeV$^{-1}$) from a single sGRB, as observed in the fluid frame, are given by
\begin{align}
& dN_p/dE_p = C_{p0} E_p^{-2} e^{-E_p/E_{p, \rm max}},\label{eq:pspec} \\
& dN_A/dE_A = C_{A0} E_A^{-2} e^{-E_A/E_{A, \rm max}},\label{eq:Aspec}
\end{align}
where $C_{p0}$ and $C_{A0}$ are the normalization constants, and 
$E_{p, \rm max}$ and $E_{A, \rm max}$ denote the maximum attainable energies of protons and $r$-process nuclei, respectively.
Without affecting the final results, we set the corresponding minimum energies in the fluid frame to be 10 and $10^3$~GeV, respectively. 

As listed in Table~\ref{tab:pars-nu}, three typical values of $\Gamma$ are considered for each emission phase: $\Gamma=300$, 500, and 800 for the PR case, and $\Gamma=20$, 30, and 50 for the EE-H, EE-L, and PE cases. The radius $R$ for each $\Gamma$ is chosen to satisfy $\tau_{A\gamma}(E_{A1})=1$. We set $\kappa_{\rm acc}=10$ for all phases, and adopt $\xi_{B/e}=0.1$ for the PR, EE-H, and EE-L cases, and $\xi_{B/e}=10$ for the PE case. The resulting 
$E_{A, \rm max}^{\rm ob}$ for all cases ranges from 100 to 1000 EeV, while the corresponding $E_{p, \rm max}^{\rm ob}$ is about 30--60 times lower (see Table~\ref{tab:pars-nu}). Based on the total energy available for each phase along with the calculated $E_{A,\rm max}$ and $E_{p, \rm max}$, we estimate $C_{p0} \approx C_{A0} \approx 4.5\times 10^{54}\Gamma^{-1}$, $ 10^{54}\Gamma^{-1}$, and $ 10^{53}\Gamma^{-1}$~GeV for the PR, EE (H and L), and PE cases, respectively.

Note that the chosen values of $R$ ($\xi_{B/e}$) for the EE (PE) phase are larger than those typically adopted in the literature.
They are selected to illustrate the potential neutrino yields for a parametric study with constraints motivated by our considerations of simple physics.

\subsection{HE neutrinos from photomeson and hadronic processes}

\begin{figure*}[htbp]  
\centering
\includegraphics[width=0.49\textwidth]{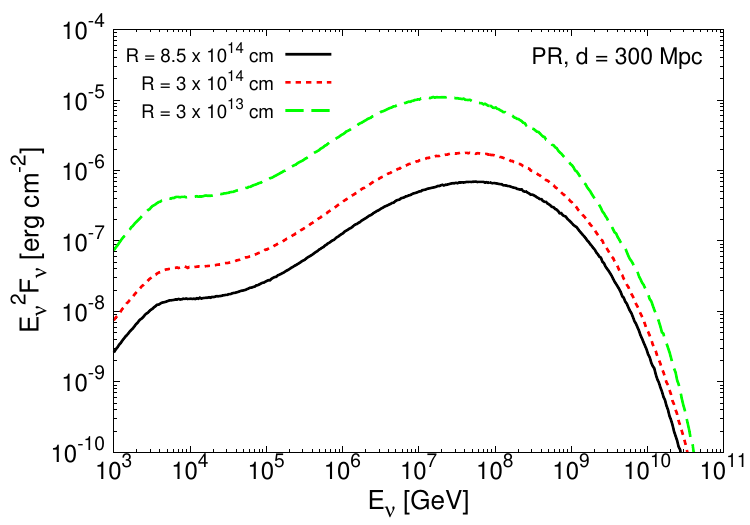} 
\includegraphics[width=0.49\textwidth]{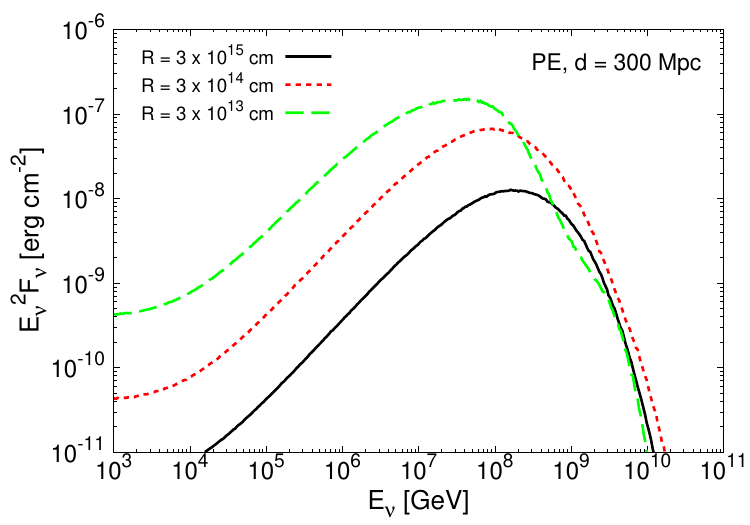} \\
\includegraphics[width=0.49\textwidth]{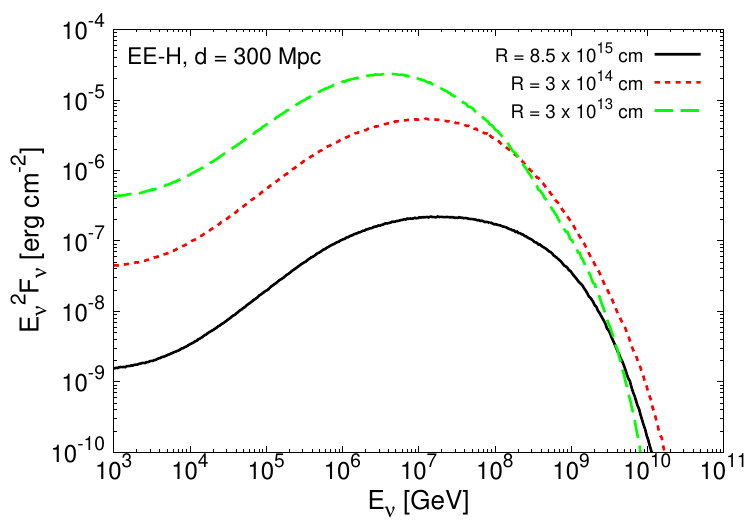} 
\includegraphics[width=0.49\textwidth]{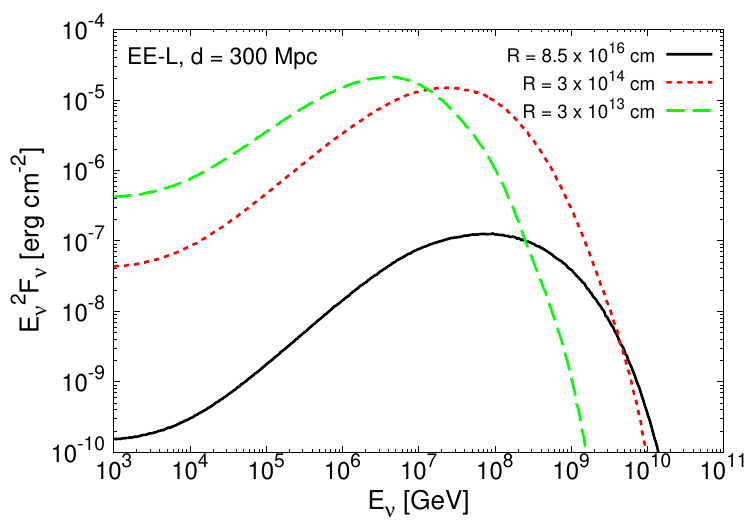}
\caption{Same as Fig.~\ref{fig:nuflx}, but for neutrino fluences originating from photonmeson and hadronic processes of protons with 
$R=3\times 10^{13}$ and $R=3\times 10^{14}$~cm, along with $\Gamma=500$ for the PR case, and $\Gamma=30$ for the EE-H, EE-L, and PE cases. For comparison, the neutrino fluences with $R$ corresponding to $\tau_{A\gamma}(E_{A1})=1$ for the same $\Gamma$ are also shown (black lines).}  
\label{fig:nuflx-R}  
\end{figure*}

We first consider the HE neutrinos produced from reactions between HE nuclei and photons or baryonic matter within the jets of sGRBs. For an HE $r$-process nucleus with energy $E_A$ in the fluid rest frame, the efficiencies of producing HE neutrinos via $A\gamma$ and $Ap$ collisions are defined as 
$f_{A\gamma}(E_A) = t^{-1}_{\rm cool, meson}(E_A) \big/t^{-1}_{\rm cool}(E_A)$ and $f_{Ap}(E_A) = t^{-1}_{\rm had}(E_A) \big/ t^{-1}_{\rm cool}(E_A)$,
where all the relevant timescales are computed as described in Sec.~\ref{sec:Emax-survival}. The resulting spectrum of meson $m$ ($m=\pi^\pm$ or $K^\pm$) produced by a single HE nucleus in the fluid frame is $f_{A\gamma}(E_A) M_{A\gamma \to m}(E_A, E_m) + f_{Ap}(E_A) M_{Ap \to m}(E_A, E_m)$, where $M_{A\gamma\to m} = dN_m^{A\gamma}/dE_m$ and $M_{Ap \to m}=dN_m^{Ap}/dE_m$ are the average meson spectra produced per $A\gamma$ and $Ap$ reaction, respectively. The meson spectrum from reactions of HE protons can be obtained similarly. To estimate the meson yields, we take the independent particle approximation, assuming $M_{A\gamma\to m}(E_A, E_m) = M_{p\gamma\to m}(E_p = E_A/A, E_m)$ and $M_{Ap\to m}(E_A, E_m) = M_{pp\to m}(E_p = E_A/A, E_m)$. As in Ref.~\cite{Guo:2022zyl}, we use SOPHIA \cite{Mucke:1999yb} and PYTHIA 8.3 \cite{Bierlich:2022pfr} to obtain the $\pi^\pm$ and $K^\pm$ yields from single $p\gamma$ and $pp$ reactions. 

We use Monte Carlo simulations to generate the HE neutrino fluences, taking into account decays of $\pi^\pm$ and $K^\pm$ (e.g., $\pi^- \to \mu^- + \bar\nu_\mu$, $K^- \to \mu^- + \bar\nu_\mu$) and subsequent decays of the produced $\mu^\pm$ (e.g., $\mu^- \to e^- + \nu_\mu + \bar\nu_e$).
For $\pi^\pm$, $K^\pm$, and $\mu^\pm$, we include the suppression factors due to synchrotron and adiabatic cooling, 
\begin{align}
f_{m, \rm sup}(E_m) =  {t^{-1}_{m, \rm d}(E_m) \over t^{-1}_{m, \rm d}(E_m) + t^{-1}_{m, \rm syn}(E_m) + t^{-1}_{\rm ad}}, 
\end{align}
where $t_{m, \rm d}(E_m) = \tau_m E_m/m_m$ is the decay time with $\tau_m$ being the mean lifetime at rest and $m_m$ the meson rest mass, and the synchrotron cooling time $t_{m, \rm syn}(E_m)$ is given by Eq.~\eqref{eq:syn} with the substitutions $m_A \to m_m$, $E_A \to E_m$, and $Z \to 1$. For $K^\pm$ decays into $\mu^\pm$, a branching ratio of $f_K \approx 0.64$ is used. The neutrino spectra from meson decays are boosted to the laboratory frame assuming isotropic distribution of $\pi^\pm$ and $K^\pm$ in the fluid rest frame.

Using the parameters in Table~\ref{tab:pars-nu}, Fig.~\ref{fig:nuflx} presents the all-flavor HE neutrino fluences (in units of $\rm erg~cm^{-2}$) arriving at a terrestrial observatory from different emission phases of a nearby sGRB at a distance of $d=300$ Mpc in terms of 
\begin{align}
E_\nu^2 F_\nu(E_\nu)\equiv  E_\nu^2 {dN_\nu \over dE_\nu}\cdot {1\over 4 \pi d^2},\label{eq:fluence}
\end{align}
where $E_\nu$ is the observed neutrino energy, and $dN_\nu/dE_\nu$ is the energy differential emissivity of HE neutrinos for the sGRB. For photomeson and hadronic processes, the HE neutrino fluences produced by the $r$-process nuclei ($^{130}_{~52}$Te) are always lower by a factor of 5--10 compared to those produced by protons. This difference is primarily due to the suppression of the $A\gamma$ and $Ap$ cross sections by nuclear shadowing and medium effects. As mentioned earlier, the HE neutrino fluences for a more realistic nuclear composition would fall between those for the pure protons and pure $r$-process nuclei shown in Fig.~\ref{fig:nuflx}.

Some features of the neutrino fluences merit discussion. The changes in the slope of $E_\nu^{2}F_\nu(E_\nu)$ at $E_\nu\sim 10^4$~GeV (see upper left panel of Fig.~\ref{fig:nuflx}) corresponds to the transition of the dominant production channel from $Ap$ ($pp$) to $A\gamma$ ($p\gamma$) reactions. The sharp drop at $E_\nu \gtrsim 10^8$ GeV is caused by the exponential falloff of the spectra of accelerated nuclei (protons) at energies above $E_{A, \rm \max}$ ($E_{p, \rm \max}$). Further suppression of the fluences occurs at even higher $E_\nu$ due to the cooling of charged mesons. Figure~\ref{fig:nuflx} shows that for the PE case, the rising part of the fluence up to the peak is approximately independent of $\Gamma$. In contrast, for the PR, EE-H, and EE-L cases, that part of the fluence decreases with increasing $\Gamma$. This difference comes about because while $\tau_{A\gamma}(E_{A1})=1$ applies to all the cases, $\tau_{A\gamma}(E_{A1}) \propto L_{\gamma, \rm iso} R^{-1}\Gamma^{-4}$ for the PE case ($\lambda<1$), but $\tau_{A\gamma}(E_{A1}) \propto L_{\gamma, \rm iso} R^{-1}\Gamma^{-2}$ for the PR, EE-H, and EE-L cases ($\lambda>1$). The fluence $E_\nu^{2}F_\nu(E_\nu)$ below the peak is proportional to $f_{A\gamma}$ ($f_{p\gamma})$, which approximately scales as $L_{\gamma, \rm iso} R^{-1} \Gamma^{-4}$ [cooling by $A\gamma$ ($p\gamma$) reactions with $\lambda<1$]. Consequently, for the PE case with $L_{\gamma, \rm iso} R^{-1}\Gamma^{-4}$ fixed by $\tau_{A\gamma}(E_{A1})=1$, that part of the fluence almost does not change with $\Gamma$. However, for the PR, EE-H, and EE-L cases with $L_{\gamma, \rm iso} R^{-1}\Gamma^{-2}$ fixed by $\tau_{A\gamma}(E_{A1})=1$, that part of the fluence approximately scales as $\Gamma^{-2}$.

\begin{figure}[htbp]  
\centering
\includegraphics[width=0.49\textwidth]{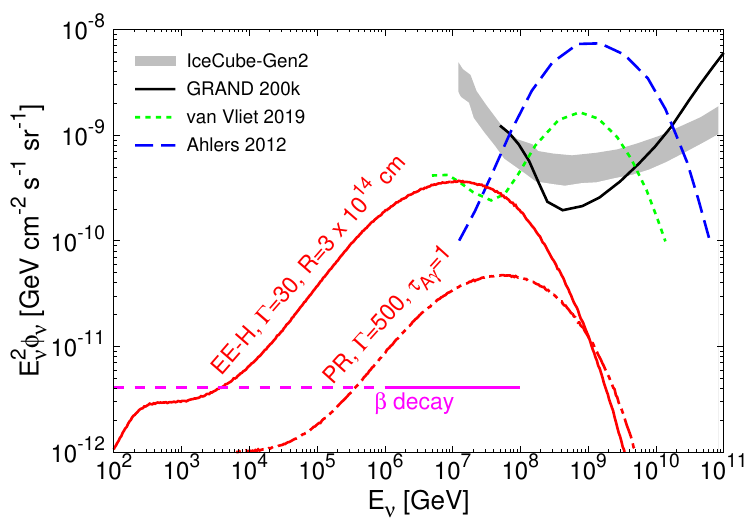}
\caption{The all-flavor diffuse neutrino flux from sGRBs in the PR and EE-H cases. Taking a pure proton composition, the neutrino flux originated from $p\gamma$ and $pp$ processes (the red lines) are obtained. We set $\Gamma=500$ and $R = 8.5\times 10^{14}$~cm for the PR case so that $\tau_{A\gamma}(E_{A1})=1$. For the EE-H case, we adopt $\Gamma=30$ and $R=3\times 10^{14}$~cm with $\tau_{A\gamma}(E_{A1})\approx 28$. The flux from $\beta$ decays of $r$-process nuclei are also shown. For comparison, two representative cosmogenic neutrino flux models are included: the model of Ahlers $et~ al.$ (assuming dominance of protons above 4 EeV) \cite{Ahlers_2012} and 
the model of van Vliet $et~al.$ (with parameters $\gamma=2.5$, $E_{\rm max}=10^{20}$ eV, $m=3.4$, and a 10\% proton fraction) \cite{vanVliet:2019nse,Meier:2024flg}. The projected sensitivities of IceCube-Gen2 (5 years) \cite{IceCube:2019pna} and GRAND200k (10 years) \cite{GRAND:2018iaj,Carpio:2020wzg} are shown for reference.} 
\label{fig:diffuse}  
\end{figure}

As mentioned in Sec.~\ref{sec:Emax-survival}, the typical values of $R$ commonly assumed in the literature, particularly for the EE phase, are smaller than those adopted here to facilitate the survival of UHE heavy nuclei. Assuming a pure proton composition, Fig.~\ref{fig:nuflx-R} shows the neutrino fluences from $p\gamma$ and $pp$ reactions for $R=3\times 10^{13}$ and $3\times 10^{14}$~cm, with $\Gamma=500$ for the PR case and $\Gamma=30$ for the EE-H, EE-L, and PE cases. For comparison, the fluences calculated with $\tau_{A\gamma}(E_{A1})=1$, using the same $\Gamma$ for each emission case, are also shown. For a given $\Gamma$, a smaller $R$ leads to a lower $E_{p, \rm max}$ ($E_{A, \rm max}$), while increasing $f_{p\gamma}$ ($\tau_{A\gamma}$). The overall neutrino fluences tend to increase as $f_{p\gamma}$ increases and the peak energies shift to lower values as $E_{p, \rm max}$ decreases (see Fig.~\ref{fig:nuflx-R}). Clearly, the neutrino fluences are limited by the requirement that the UHE heavy nuclei survive with sufficient probability \cite{Murase:2010gj}. In other words, enhanced neutrino yields for smaller $R$ come at the cost of suppressing UHE $r$-process nuclei.
Using the effective area of IceCube averaged over the declination angle \cite{IceCube:2014vjc} and the $\sim 10^{2/3}$ times larger effective area of IceCube-Gen2 \cite{IceCube:2014vjc}, we estimate that IceCube (IceCube-Gen2) would observe $\sim 4\times 10^{-4}$ ($\sim 2\times 10^{-3}$) upgoing muon events induced by HE neutrinos emitted during the PR phase of an sGRB at 300 Mpc when $\tau_{A\gamma}(E_{A1})=1$ is imposed. With smaller $R$, the expected event number at IceCube-Gen2 could reach ${\cal O}$(0.1) for the PR and EE phases.

\subsection{HE neutrinos from $\beta$ decay of radioactive $r$-process nuclei}

In addition to HE neutrinos produced via photomeson and hadronic interactions, another potential production channel is the $\beta$ decay of radioactive $r$-process nuclei that have been accelerated within relativistic jets. The $r$-process and the launching of outflows and relativistic jets occur on a timescale of ${\cal O}(1)$ s, which suggests that only nuclei with $\beta$ decay lifetimes $\gtrsim {\cal O}(1)$~s can be accelerated before undergoing significant decay. As the decay lifetimes of accelerated nuclei are Lorentz boosted and can easily exceed the typical dynamical timescale of the jets, HE neutrinos can be produced only if these accelerated nuclei escape from the jets efficiently and then undergo $\beta$ decay. Otherwise, adiabatic cooling of these nuclei inside the jets will suppress the production of HE neutrinos from their decay.

Reference~\cite{Zhang:2024sjp} suggested that UHE $r$-process nuclei with an energy generation rate density of $\sim 10^{43}$~erg~Mpc$^{-3}$~yr$^{-1}$ might be needed to better account for the observed UHECRs above 100 EeV. Given the intrinsic occurrence rate of $\sim 300$~Gpc$^{-3}$~yr$^{-1}$ for sGRBs \cite{Wanderman:2014eza,LIGOScientific:2016hpm}, the energy in UHE $r$-process nuclei is $\sim 3\times 10^{49}$~erg per sGRB. For a beaming correction factor of $f_b \sim 0.01$ \cite{Berger:2013jza}, the corresponding isotropic energy in UHE heavy nuclei is $\sim 3\times 10^{51}$~erg.\footnote{This normalization differs from that for calculating HE neutrino fluences produced by photomeson and hadronic processes within the jets during all emission phases except for the PE phase. In both cases, we try to use the normalization that is appropriate for our approach to estimate the neutrino fluences. For production by photomeson and hadronic processes, the proper normalization for the total energy of accelerated nuclei is related to the observed photon luminosity because both quantities can be viewed as properties of the jets.} As this estimate corresponds to the energy of accelerated nuclei that successfully escape from the source, it is closely related to the production of HE neutrinos from the $\beta$ decay of these nuclei. So we use it to calculate the fluxes across all emission phases. The isotropic energy spectrum (in GeV$^{-1}$) of heavy nuclei escaping from the jets is assumed to follow a power-law distribution, 
\begin{align}
{dN_A \over dE_A} = C_{A} E_A^{-\gamma_A},
\end{align}
where the normalization constant $C_A$ is determined by the total energy budget. We adopt a minimum energy of 10 EeV for the escaping nuclei, noting that the precise value depends on their escape efficiency. The maximum energy is assumed to lie in the range of $10^2$--$10^3$ EeV (see Table~\ref{tab:pars-nu}), and for simplicity, we fix it at $10^3$ EeV in our estimates. For a spectral index $\gamma_A=2$, the normalization constant is approximately 
$C_A \approx 4\times 10^{53}$~GeV.

On average, each radioactive $r$-process nucleus undergoes a decay chain that produces $\sim 10$ neutrinos, each carrying an energy of $E_\nu \sim 10^{-4} E_A$. Consequently, the isotropic neutrino spectrum emitted from a single sGRB is given by 
\begin{align}
E_\nu^2{dN_\nu \over dE_\nu} &\approx  10^{-3} \Big(E_A^2 {dN_A \over dE_A}\Big) \Bigg|_{E_A=10^4 E_\nu} \nonumber \\ & \approx 4\times 10^{50}~{\rm GeV}. 
\end{align}
Note that the typical neutrino energies are expected to lie in the range of $1$--100~PeV, corresponding to the $\beta$ decays of $r$-process nuclei with energies between 10 and $10^3$ EeV. However, the neutrino spectrum could extend to lower energies if nuclei with lower energies are also able to escape from the source. The $\beta$ decay neutrino fluence arriving at a terrestrial observatory from a nearby sGRB at 300 Mpc is shown in Fig.~\ref{fig:nuflx}.
Due to our choice of normalization, this fluence is the same for all emission phases. It can be seen from Fig.~\ref{fig:nuflx} that this fluence is comparable to or exceeds those produced by photomeson and hadronic interactions of pure $r$-process nuclei. Notably, for the PE phase, $\beta$ decay can even dominate photomeson and hadronic interactions of pure protons in producing HE neutrinos.

\subsection{Diffuse HE neutrino flux from sGRBs}

The diffuse neutrino flux (in units of $\rm GeV~cm^{-2}~ s^{-1}~sr^{-1}$) from sGRBs throughout cosmic history can be estimated as \cite{Murase:2013ffa,Guo:2019ljp,Guo:2022zyl} 
\begin{align}
E_\nu^2 \phi_\nu \sim {c R_{\rm sGRB} f_z \over 4\pi H_0} 
E_\nu^2 {dN_\nu \over dE_\nu},  
\end{align}
where $H_0 \approx 70~{\rm km~s^{-1}~Mpc^{-1}}$ is the Hubble constant, $R_{\rm sGRB} \sim 3~\rm Gpc^{-3}~yr^{-1}$ is the local rate of sGRBs (including the beaming effect) \cite{Guetta:2005bb,Wanderman:2014eza}, and $f_z \sim 3$ accounts for evolution of the sGRB rate and neutrino energy with redshift \cite{Waxman:1998yy}.

Figure~\ref{fig:diffuse} displays the all-flavor diffuse neutrino flux from $p\gamma$ and $pp$ processes (assuming a pure proton composition of the accelerated nuclei) in the PR and EE-H cases. For the PR case, we adopt $\Gamma=500$ and $R=8.5\times 10^{14}$~cm, corresponding to $\tau_{A\gamma}(E_{A1})=1$. This result serves as a representative case favorable for the production of UHE $r$-process nuclei. For the EE-H case, we take $\Gamma=30$ and $R=3\times 10^{14}$ cm, which leads to $\tau_{A\gamma}(E_{A1})\approx 28$. While this parameter set is not suitable for the survival of UHE heavy nuclei, it results in a significantly larger neutrino flux (see Fig.~\ref{fig:diffuse}). The fluxes in both cases are consistent with the current upper limits from IceCube stacking analyses \cite{IceCube:2017amx,Abbasi:2022whi}, which are at the level of $10^{-9}~{\rm GeV~cm^{-2}~s^{-1}~sr^{-1}}$ for the energy range of 1--100 PeV. The lower diffuse flux from $\beta$ decay neutrinos is also shown in Fig.~\ref{fig:diffuse}, as well as two representative cosmogenic flux models from Ahlers {\it et al}. \cite{Ahlers_2012} and van Vliet {\it et al}. \cite{vanVliet:2019nse,Meier:2024flg}. Cosmogenic neutrinos are produced by the interactions of UHECRs with the cosmic microwave background and the extragalactic background light. Their diffuse flux has relatively large theoretical uncertainties (see, e.g., \cite{Wittkowski:2018giy}). As can be seen from Fig.~\ref{fig:diffuse}, the flux from the EE phase, using the chosen parameter set, lies slightly below the projected detection sensitivities of IceCube-Gen2 (5 years) and GRAND200k (10 years) at $E_\nu \sim 10^8$ GeV. However, at $E_\nu\sim 10^7$ GeV, this flux can exceed the cosmogenic neutrino flux and approach the IceCube upper limit for GRBs, indicating a promising opportunity for detection in the near future \cite{KM3Net:2016zxf,IceCube-Gen2:2020qha,TRIDENT:2022hql}.

\section{Discussion and conclusions}
\label{sec:summary}

We have conducted a detailed investigation of the jet conditions during the PR, EE, and PE emission phases of sGRBs that could support the production and preservation of UHE $r$-process nuclei. Specifically, we consider two requirements for sGRB jets to significantly contribute to the observed CR events above 100 EeV. First, $r$-process nuclei must be accelerated to energies of at least $100$ EeV. Second, UHE heavy nuclei must survive efficiently within the jets, avoiding destruction by the intense photon background. For jets to meet these criteria, we have derived constraints on the relevant parameters, including the bulk Lorentz factor ($\Gamma$), the dissipation radius ($R$), and the energy partition parameter ($\xi_{B/e}$). Our findings, combined with predictions from existing models, indicate that the PR phase of sGRBs with $\Gamma \gtrsim 400$--500 provides the most favorable conditions for contributing UHE $r$-process nuclei. While not entirely ruled out, the EE phase is generally less favorable due to their typically smaller dissipation radii, which hinder the survival of UHE heavy nuclei. For the PE phase, achieving sufficiently high energies for heavy nuclei would require a stronger magnetic field strength or a larger $\xi_{B/e}$ than typically anticipated. 

The potential link between sGRBs and UHECRs can also be examined from an energy budget perspective. To match the observed UHECR data above 1 EeV, the required energy generation rate density of UHECRs is estimated to be $\sim 5 \times 10^{44}$~erg~Mpc$^{-3}$~yr$^{-1}$, assuming a mixed composition of conventional nuclei \cite{Auger:2016use,PierreAuger:2020kuy,Zhang:2024sjp}. This estimate corresponds to an average energy of $\sim 10^{51}$~erg in UHE conventional nuclei per sGRB, which is much higher than the energy ($\sim 3\times 10^{49}$~erg) needed in UHE $r$-process nuclei to explain the CR component above 100 EeV \cite{Zhang:2024sjp}. Assuming a beaming correction factor of $\sim 0.01$, the total available energies in HE nuclei per sGRB are about $10^{51}$, $3\times 10^{50}$, and $3\times 10^{49}$~erg for the PR, EE, and PE phases in our study, respectively. Therefore, sGRBs are energetically capable of producing the UHE $r$-process nuclei above 100 EeV during the PR and EE phases, while the PE phase may contribute marginally. If the particle escape efficiency is high, the PR phase could also significantly contribute to the observed UHECRs below 100 EeV (dominated by conventional nuclei). As the available energy budget in the PR phase is much larger than the amount needed to explain the $r$-process component of UHECRs, it is likely that the accelerated nuclei in this phase are dominated by conventional ones, in order to avoid overproducing the UHE $r$-process nuclei. In contrast, if the PE phase is primarily responsible for generating the UHE $r$-process nuclei, due to its limited energy budget, the accelerated nuclei in this phase would likely be dominated by $r$-process nuclei.

Using jet parameters for sGRBs to contribute significantly to the CR events above 100 EeV, we have calculated the resulting HE neutrino fluences from a nearby sGRB during different emission phases, assuming two limiting compositions of the accelerated nuclei (either pure protons or solely $r$-process nuclei). Our results indicate that HE neutrino production is most prominent during the PR phase when accelerated protons dominate. This is primarily due to the higher total energy available during this phase, along with the suppression of the $A\gamma$ cross section compared to the $p\gamma$ one.
Notably, the HE neutrino flux, primarily generated through photomeson production, is limited by the requirement to prevent the destruction of $r$-process nuclei through reactions with photons.
Relaxing this constraint would increase the predicted neutrino fluxes but would simultaneously suppress the contribution to the most energetic CRs. As the PR phase is more favorable for the production of UHE heavy nuclei compared to the other emission phases, constraints on HE neutrino production are particularly relevant for this phase. While the EE phase is less likely to contribute to the most energetic UHECRs, it may serve as a more promising source of neutrinos in the 1--100 PeV range than the PR phase \cite{Kimura:2017kan}. The detection or nondetection of HE neutrinos from a nearby sGRB during each emission phase could provide critical insights into the potential link between sGRBs and the most energetic CRs, which are assumed here to be dominated by $r$-process nuclei.

In addition to HE neutrinos produced by photomeson and hadronic interactions, we have also studied those from the $\beta$ decay of accelerated $r$-process nuclei. Under simplified assumptions, we find that their flux could be comparable to those from photomeson reactions.
Further, we have estimated the diffuse HE neutrino flux from photomeson and hadronic interactions of protons during the PR (EE) phase, adopting parameters that favor (disfavor) contribution to UHE $r$-process nuclei. While the flux from the PR phase remains well below the detection sensitivities, the higher diffuse flux in the energy range of 1--100 PeV from the EE phase could be detected by IceCube and the next-generation neutrino observatories.         

In this study, we have adopted typical luminosities for sGRBs. Future work should explore the distribution of luminosities to provide a more complete study. In addition, we have assumed that some $r$-process nuclei synthesized in the nonrelativistic ejecta are somehow mixed into the relativistic jets.
The inferred total energy of UHE $r$-process nuclei per sGRB corresponds to a tiny amount ($\sim 10^{-14}~M_\odot$)
of $r$-process nuclei accelerated above 100 EeV. Considering a power law spectrum for the accelerated nuclei, which are only a fraction of the nuclei in the jets, it may require up to $\sim 10^{-6}~M_{\odot}$ of $r$-process nuclei to be mixed into the relativistic jets
to account for the observed UHECRs above 100 EeV. It remains to be explored how this mixing can be achieved. While this work primarily focuses on BNSMs and sGRBs, similar analyses and conclusions could be extended to collapsars and lGRBs in a straightforward manner.

\begin{acknowledgments}
This work was supported in part by the National Natural Science Foundation of China (No.~12205258), Guangdong Basic and Applied Basic Research Foundation (No.~2025A1515011082), the ``CUG Scholar" Scientific
Research Funds at China University of Geosciences (Wuhan) [No.~2021108 (G.G.)], the US Department of Energy [DE-FG02-87ER40328 (Y.Z.Q.)], the National Science and Technology Council (No.~111-2628-M-001-003-MY4), the Academia Sinica (No.~AS-IV-114-M04), and the Physics Division of the National Center for Theoretical Sciences, Taiwan (M.R.W.).    
\end{acknowledgments}

\appendix
\section{Constraints on the parameters for the EE-H, EE-L, and PE cases}\label{app:constraints}
\renewcommand{\thefigure}{A\arabic{figure}}
\setcounter{figure}{0}

\begin{figure*}[htbp]  
\centering
\includegraphics[width=0.49\textwidth]{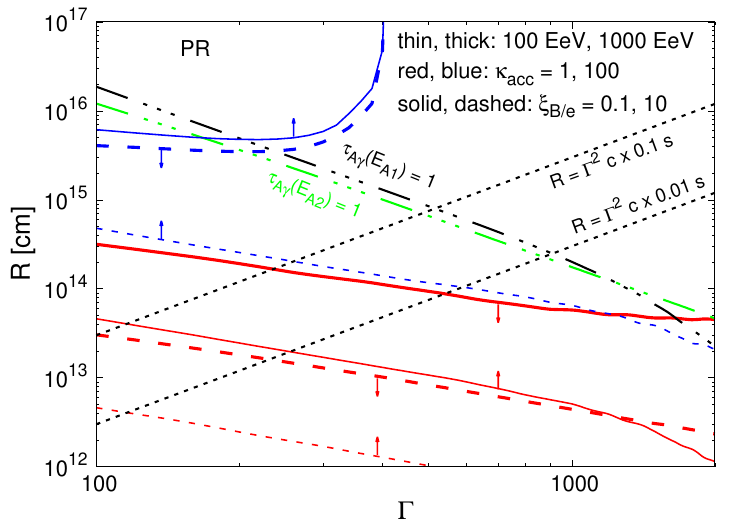} 
\includegraphics[width=0.49\textwidth]{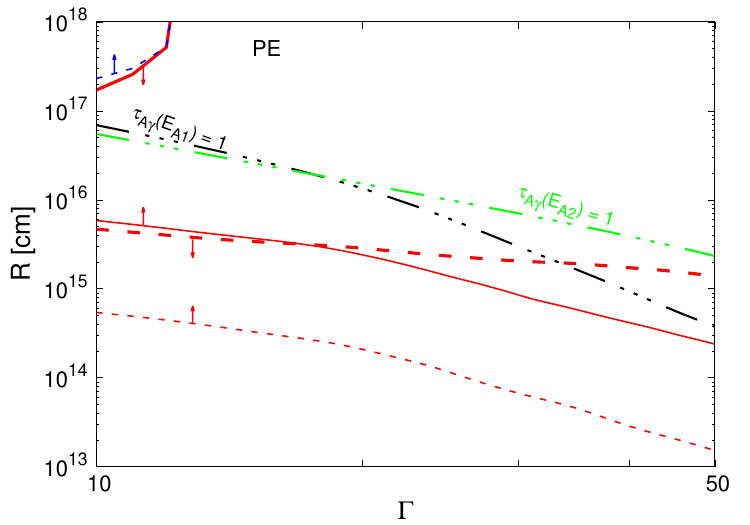} \\
\includegraphics[width=0.49\textwidth]{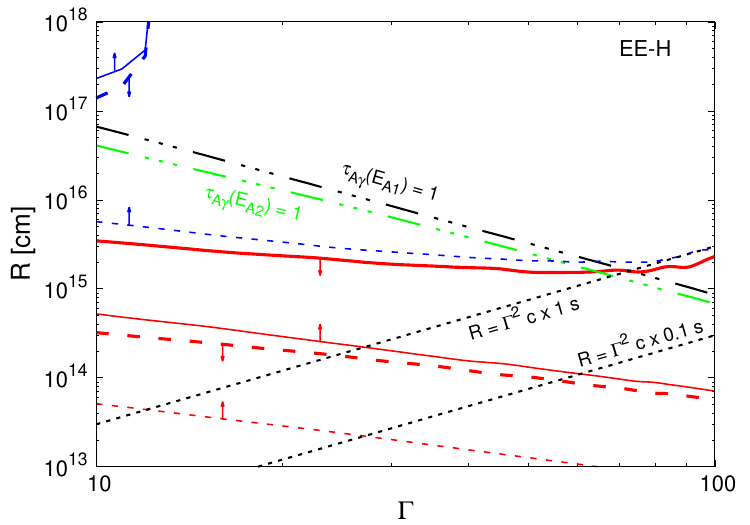} 
\includegraphics[width=0.49\textwidth]{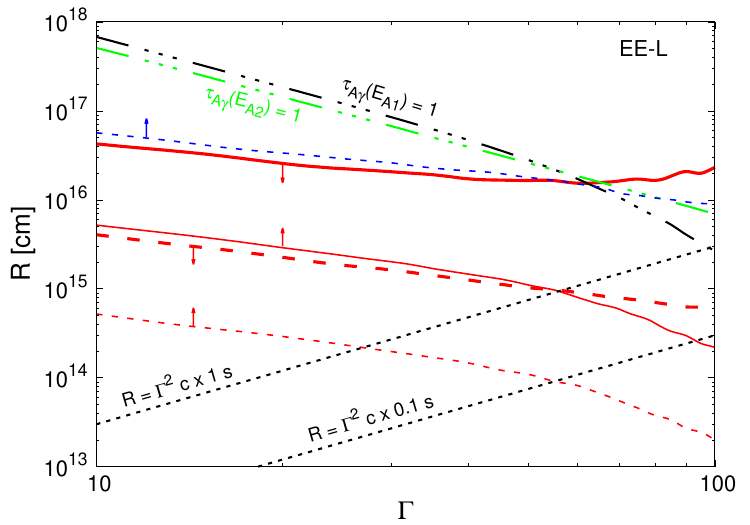}
\caption{Contours in the $\Gamma$-$R$ plane corresponding to $\tau_{A\gamma}(E_{A1})=1$ (black lines), $\tau_{A\gamma}(E_{A2})=1$ (green lines), and $E_{A, \rm max}^{\rm ob}=100$ and 1000 EeV (red and blue lines) for the PR (upper left), PE (upper right), EE-H (lower left), and EE-L (lower right) cases. The heavy nuclei are assumed to be $^{130}_{~52}$Te. To derive the contours for $E^{\rm ob}_{A, \rm max}$, we have taken $\kappa_{\rm acc}=1$ and 100, and $\xi_{B/e}=0.1$ and 100. For the PR, EE-H, and EE-L cases, the expected radii of internal shocks are also indicated (black dashed lines) with $R \approx \Gamma^2 ct_v$, where $t_v$ is the timescale of variability. For the PR case, we set the minimum and maximum $t_v$ to 0.01 and 0.1 s, respectively, while for the EE case, $t_v$ ranges from 1 to 10 s. See text for details.} 
\label{fig:LFR_range_all}  
\end{figure*}

Figure~\ref{fig:LFR_range_all} presents the constraints on the $\Gamma$--$R$ parameter space for all emission cases, considering different combinations of $\xi_{B/e} = 0.1$, 10 and $\kappa_{\rm acc} = 1$, 100. The upper left panel summarizes the PR case shown in Fig.~\ref{fig:LFR_range}. Here we focus on the EE-H, EE-L, and PE cases.

As discussed for the PR case, $\tau_{A\gamma}\propto L_{\gamma,\rm iso}R^{-1} \Gamma^{-2}$ is approximately independent of $E_A$ due to cooling by meson production with $\lambda=2E_A^{\rm ob} E_{\gamma b}^{\rm ob}/(\Gamma^2 m_A\epsilon_\Delta)>1$. For the EE-L and PE cases with lower $E_{\gamma b}^{\rm ob}$ (see Table~\ref{tab:pars}), $\lambda<1$ occurs for $E_{A1}^{\rm ob}=100$~EeV at large $\Gamma$ and $\tau_{A\gamma} \propto L_{\gamma,\rm iso}R^{-1} \Gamma^{-4}$ [see the contour of $\tau_{A\gamma}(E_{A1})=1$ following $R\propto\Gamma^{-4}$ in the upper right panel of Fig.~\ref{fig:LFR_range_all}]. The contours of $\tau_{A\gamma}(E_{A1})=1$ (dash-dotted lines) in Fig.~\ref{fig:LFR_range_all} indicate that the survival of UHE heavy nuclei requires $R$ exceeding $\sim 10^{15}$~cm for the EE-H and EE-L cases, and $\sim 10^{14}$~cm for the PE case.
Notably, the required radii for the EE phase exceed the values typically adopted in the literature ($R \sim 10^{13}$--$10^{15}$ cm) \cite{Metzger:2007cd,Kimura:2017kan}. Further, when applying the IS model to the EE phase, the dissipation radii with $t_v = 1$--10 s (indicated by the black dotted lines) fall below the contours of $\tau_{A\gamma}(E_{A1})=1$ unless $\Gamma$ is near or above 100, suggesting that UHE heavy nuclei are unlikely to survive under typical EE conditions.

The red and blue lines in Fig.~\ref{fig:LFR_range_all} represent the contours of $E^{\rm ob}_{A, \rm max}=100$ and 1000 EeV. As discussed for the PR case, $E_{A, \rm max}^{\rm ob}\propto \xi_{B/e}^{1/2}\kappa_{\rm acc}^{-1}\Gamma R$ when cooling by meson production with $\lambda>1$ dominates or $E_{A, \rm max}^{\rm ob}\propto \xi_{B/e}^{1/2}\kappa_{\rm acc}^{-1}\Gamma^{-1}$ when adiabatic cooling dominates. When cooling by meson production with $\lambda<1$ dominates, $E_{A, \rm max}^{\rm ob}\propto \xi_{B/e}^{1/4}\kappa_{\rm acc}^{-1/2}\Gamma^{3/2} R^{1/2}$ (see the contours of $E_{A, \rm max}^{\rm ob}=100$~EeV following $R\propto\Gamma^{-3}$ in the right panels of Fig.~\ref{fig:LFR_range_all}.)
As $\xi_{B/e}$ increases and $\kappa_{\rm acc}$ decreases, heavy nuclei are easier to accelerate, causing the $E^{\rm ob}_{A, \rm max}$ contours to shift toward lower $R$ for a given $\Gamma$. Similar to the PR case, the parameter set $\kappa_{\rm acc}=1$ and $\xi_{B/e}=10$ is ruled out for the EE cases because the corresponding $E^{\rm ob}_{A, \rm max}$ contours fall below the contour of $\tau_{A\gamma}(E_{A1})=1$. On the other hand, for the PE phase, the combination of $\kappa_{\rm acc}=100$ and $\xi_{B/e}=0.1$ leads to insufficient acceleration and the required $E^{\rm ob}_{A, \rm max}$ cannot be reached for the adopted range of $\Gamma$.

It is interesting to note that because $\tau_{A\gamma}(E_{A2})>\tau_{A\gamma}(E_{A1})$ at $\Gamma \gtrsim 20$ for the PE case (see the upper right panel of Fig.~\ref{fig:LFR_range_all}), the inferred spectra of UHECRs at 100--1000~EeV can also be explained by the requirements:
\begin{align}
E_{A, \rm max}^{\rm ob} > E_{A2}^{\rm ob}~{\rm and}~\tau_{A\gamma}(E_{A1})<1<\tau_{A\gamma}(E_{A2}), \label{eq:req-2} 
\end{align}
for which heavy nuclei are accelerated to above 1000 EeV, but are subsequently destroyed by $A\gamma$ reactions while those nuclei accelerated to 100--1000~EeV survive with sufficiently high probability. As Eq.~\eqref{eq:req-2} does not impose an upper bound on $E_{A, \rm max}^{\rm ob}$, including the corresponding scenario effectively removes the upper limit on $\xi_{B/e}$ as a function of $\kappa_{\rm acc}$ shown in Fig.~\ref{fig:kap-Xibe} for the PE phase.

\end{document}